\documentclass[letter,12pt,titlepage]{article}
\usepackage[utf8]{inputenc}
\usepackage[margin=1in]{geometry}
\usepackage[nodisplayskipstretch]{setspace}

\usepackage{mathtools,amssymb,microtype,bm,dsfont,graphicx,xcolor,soul,placeins,booktabs,multicol,multirow,rotating}
\usepackage{adjustbox,array,bbm}

\usepackage{titling}\thanksmarkseries{fnsymbol}
\pretitle{\begin{center}\Large}
\preauthor{\begin{center}\normalsize\lineskip 1em \begin{tabular}[t]{c}}
\predate{\begin{center}\normalsize}

\usepackage{marginnote}
\newcounter{todo}

\usepackage{newtxtext,newtxmath}

\pretitle{\begin{center}\bfseries}
\posttitle{\par\end{center}\vskip 2em}
\preauthor{\begin{center}\normalfont\lineskip 0.5em \begin{tabular}[t]{c}}
\postauthor{\end{tabular}\par\end{center}}
\predate{\begin{center}\normalfont}
\postdate{\par\end{center}}

\usepackage[explicit]{titlesec}
\titleformat{\section}[hang]{}{}{0em}{\centering\normalfont\normalsize\bfseries\MakeUppercase{#1}}
\titlespacing*{\section}{0pt}{1em}{*0}
\titleformat{\subsection}[hang]{}{}{0em}{\normalfont\normalsize\bfseries#1}
\titlespacing*{\subsection}{0pt}{1em}{*0}
\titleformat{\subsubsection}[runin]{}{}{0em}{\normalfont\normalsize\bfseries\itshape#1.}
\titlespacing*{\subsubsection}{\parindent}{0pt}{1.5ex plus .2ex}

\usepackage[labelfont=bf]{caption}

\usepackage[natbibapa]{aom}
\bibpunct[, ]{(}{)}{,}{a}{}{,}
\def\bibhang{24pt}  
 \def\bibsep{\smallskipamount}
\setcitestyle{notesep={:\,}}\newcommand{\p}[1]{#1}\newcommand{\pp}[1]{#1}

\let\oldappendix\appendix
\renewcommand{\appendix}{\oldappendix\titleformat{\section}[block]{\normalfont\Large\bfseries}{Appendix \thesection:}{0.5em}{##1}\setcounter{figure}{0}\numberwithin{figure}{section}\setcounter{table}{0}\numberwithin{table}{section}}

\usepackage[breaklinks,hidelinks]{hyperref}

\renewcommand{\vec}[1]{\bm{\mathbf{#1}}}
\newcommand{\one}{\mathbbm{1}}

\newcommand{\TITLE}{Search Heuristics Under Multiple Objectives:\\The Case of Corporate Social Responsibility}
\title{\TITLE}
\author{Daniel Albert \\ LeBow College of Business \\ Drexel University \\ Philadelphia, PA 19104 \\ \texttt{daniel.albert@drexel.edu} \and Felipe A. Csaszar \\ Ross School of Business \\ University of Michigan \\ Ann Arbor, MI 48109 \\ \texttt{fcsaszar@umich.edu}}
\date{{\vskip 1em}June 12, 2026{\vskip 2em}Forthcoming in \emph{Academy of Management Review}.}

\begin{document}
\begin{onehalfspace}\maketitle
\begin{abstract}
Organizations increasingly pursue multiple objectives, yet we know little about how boundedly rational firms search for better strategies when performance is multidimensional.  We use corporate social responsibility (CSR)---pursuing social and financial outcomes simultaneously---as a motivating context.  We distinguish multi-objective \emph{decision making} (choosing among known alternatives) from multi-objective \emph{search} (discovering alternatives through path-dependent local moves), and argue that preferences matter through the heuristics that implement them during search.  In a dual-landscape NK simulation, we compare five prominent heuristics: \emph{Maximize} (improve financial performance only), \emph{Combine} (raise the sum of financial and social performance), \emph{Alternate} (pursue one goal until stuck, then switch), \emph{Penalize} (maximize financial performance while deducting shortfalls below a social threshold), and \emph{Satisfice} (prioritize financial performance only after meeting a social threshold).  Across varying complexity, goal correlation, and thresholds, these heuristics produce systematically different trajectories and joint outcomes.  Importantly, by escaping local financial optima, \emph{Alternate} often discovers ``oblique'' strategies that match or exceed the financial performance achieved by profit-only local search while also improving social performance.  We conclude that implementing the same multi-goal preferences through different search heuristics steers firms toward different regions of the social--financial frontier, shaping both the compromises they reach and the opportunities they discover.

\end{abstract}
\end{onehalfspace}

\section{Introduction}

\subsection{Searching Under Multiple Goals: The Behavioral Challenge}

Strategic decision-making is often presented as a problem of ranking alternatives on a single performance metric---such as return on investment or net present value---and selecting the highest-scoring option.  Yet many strategic choices are inherently multidimensional: managers must evaluate alternatives that simultaneously affect cost and quality, speed and safety in product development, mission and margin in public and nonprofit services, and (increasingly) social and financial outcomes in corporate strategy.  In such settings, comparing alternatives is not simply a matter of asking whether one scalar is greater than another; it requires comparing vectors.  For example, consider two projects: Project $A$ yields \$100 million in profits and two units of social value, while Project $B$ yields \$200 million but only one unit of social value.  Which is better?  Unlike scalars, vectors lack a total order \citep{Halm74}.  Consequently, even defining what it means to move ``up'' or ``down'' in performance becomes ambiguous---and this ambiguity is not a peripheral complication but a fundamental behavioral challenge whenever organizations pursue multiple goals.

This challenge is especially salient in corporate social responsibility (CSR), where firms aim to ``do well and do good'' by improving both corporate financial performance (CFP) and corporate social performance (CSP).  These goals are conceptually distinct and, empirically, only weakly correlated on average \citep{Vish20}.  Meta-analyses summarize hundreds of studies reporting correlations between CSP and CFP that range from modestly positive to strongly negative depending on context \citep{Marg03,Orli03,Vish20}.  In some industries, technological and contracting conditions produce systematic trade-offs (e.g., cost--quality or environmental trade-offs), making it difficult to raise one dimension without lowering the other \citep{Hart97b}.  In other settings, stakeholder reactions and shared-value opportunities can create positive associations between CSP and CFP \citep{Flam15,Doro17}.  Thus, the relevant strategic problem is not merely whether CSP and CFP are aligned or in conflict, but how organizations pursue improvements in environments where the relationship between objectives varies.

Most research addressing such multidimensionality has focused on \emph{decision-making}: given a set of known alternatives, how should actors evaluate and select among them under multiple criteria?  The multi-objective decision-making literature formalizes this problem using concepts such as domination and the efficient frontier \citep{Keen76,Ehrg05,Emme18}.  However, this one-shot framing treats the set of alternatives as given.  By contrast, the behavioral theory of the firm emphasizes \emph{search}: under bounded rationality, alternatives are not given but must be discovered over time through local, iterative exploration \citep{Simo55,Cyer63,Levi97}.  Search is therefore logically prior to choice and path dependent: the alternative adopted today becomes the starting point for discovering what is feasible tomorrow.  This distinction is consequential in a multi-goal setting because the rules used to implement preferences are not only decision criteria applied to known options; they are \emph{search heuristics} that steer which options the firm ever discovers.  Yet we still lack a clear understanding of how different multi-goal search heuristics shape search trajectories and, ultimately, the joint social and financial outcomes organizations achieve.

\subsection{Our approach and contribution}

We address this gap by studying search under multiple objectives.  Using CSR as a motivating and empirically rich context, we ask: \emph{How do different search heuristics affect the social and financial outcomes organizations achieve when performance is multidimensional?}  To answer this question, we model organizations as boundedly rational searchers on rugged performance landscapes and examine how different rules for evaluating nearby alternatives generate different search trajectories and, consequently, different outcomes.  We focus on five heuristics drawn from the CSR and multiple-goals literatures.  \emph{Maximize} represents a profit-first rule that ignores social performance \citep{Frie70,Jens02}.  \emph{Combine} aggregates social and financial outcomes into a single composite metric, reflecting additive or utilitarian-like evaluation \citep{Bent1789,Mill1848}.  \emph{Alternate} sequentially shifts attention between objectives, consistent with the behavioral argument that organizations often attend to one goal at a time \citep{Cyer63,Grev08}.  \emph{Penalize} maximizes financial performance while imposing explicit penalties for falling below a social threshold \citep{McWi01}.  \emph{Satisfice} prioritizes financial performance only after meeting a minimum social aspiration, reflecting problemistic search and threshold-based evaluation \citep{Marc58,Cyer63}.  These heuristics are not merely different preference profiles; some encode similar preferences but implement them differently (e.g., Combine and Alternate can both reflect joint concern for CSP and CFP), allowing us to isolate how \emph{implementing} preferences via different search heuristics changes what firms discover and achieve.

We implement these heuristics in an NK simulation model extended to dual performance landscapes \citep{Levi97,Adne14}.  Each strategy is a vector of interdependent decisions that jointly determines both financial and social performance, and we vary the correlation between the two performance landscapes to represent environments with trade-offs, independence, or complementarities.  This approach lets us examine how the relative effectiveness of heuristics depends on both internal search processes and external performance structure---a multi-goal instantiation of Simon's scissors \citep{Simo90b}.

Our analysis makes four contributions.  First, we advance behavioral strategy by showing that, under multiple objectives, the heuristics that implement preferences are consequential not only for \emph{choice} among known alternatives but also for \emph{search} itself: they shape which alternatives organizations discover through path-dependent exploration, clarifying why the search/decision-making distinction matters in multi-goal settings \citep{Keen76,Cyer63}.  Second, we provide a systematic comparative analysis of five prominent multi-goal heuristics---Maximize, Combine, Alternate, Penalize, and Satisfice---and show how their relative effectiveness depends on the structure of the environment (e.g., whether social and financial outcomes are in tension or mutually reinforcing).  Third, we uncover an important implication for CSR: alternating attention between social and financial goals can, in many conditions, improve social outcomes \emph{without sacrificing} financial performance---and can even lead firms to \emph{higher} financial performance than a narrow focus on profit maximization.  By shifting between objectives, firms can escape a local financial optimum at which a profit-maximizing firm settles, and thereby potentially discover strategies that perform better on both dimensions.  This is consistent with the logic of ``oblique strategies'' \citep{Eno01,Kay10}, that is, the notion that high-performing solutions are discovered through detours rather than a persistent focus.  This finding also challenges strong versions of the shareholder-primacy argument \citep{Frie70,Jens02}.  Fourth, we distill the modeled heuristics into a parsimonious ``grammar'' for describing and contrasting multi-objective search heuristics, providing a template that can be extended to other multi-goal contexts beyond CSR.

The remainder of the paper is organized as follows.  Section~\ref{sec:theory} develops the theoretical foundation by (i) formalizing the challenge of multidimensional performance, (ii) summarizing multi-objective decision-making concepts, (iii) shifting to the behavioral logic of multi-objective search, and (iv) motivating the five heuristics.  Section~\ref{sec:model} presents the dual-landscape NK model and our simulation design.  Section~\ref{sec:results} reports comparative results across environments, including conditions under which multi-objective heuristics match or outperform Maximize.  Section~\ref{sec:discussion} discusses implications for behavioral strategy and CSR, including the practical relevance of heuristic choice when organizations face multiple objectives.

\section{Theoretical Background}\label{sec:theory}

This section establishes the theoretical foundation for our analysis of how firms search for strategies that satisfy multiple objectives.  We focus on the specific, yet broadly relevant, context of corporate social responsibility (CSR), where firms face the dual imperatives of financial and social performance.  We proceed in four parts.  First, we frame the core challenge: the inherent ambiguity of comparing strategic alternatives when performance is multidimensional.  Second, we introduce key ideas from the literature on multi-objective decision making, which provides a formal framework for comparing and selecting among alternatives when objectives are multidimensional.  Third, we discuss the role of multi-objective search within the behavioral theory of the firm, emphasizing the generation and discovery of alternative strategies under bounded rationality.  Finally, we distill five search heuristics from the CSR and multiple-goals literatures, forming the basis for our subsequent model.

\subsection{The Challenge of Multidimensional Performance}

A fundamental difficulty in managing for multiple objectives stems from the fact that performance is not a single, scalar value, but rather a vector with multiple dimensions (consistent with Miller et al.'s \citeyear{Mill13} view of performance as a domain of ``separate constructs'').  Multiple objectives reflect organizational members' (and potentially external stakeholders') diverse, and sometimes conflicting, interests, requiring organizations to consider multiple performance dimensions simultaneously---financial, social, environmental, and others.  This multidimensionality is central to classic and contemporary work on multiple goals in organizations \citep{Cyer63,Grev08,Gaba19,Grev20}, and it is increasingly salient in strategy research emphasizing that firms must simultaneously manage multiple performance dimensions (e.g., mission vs.\ margin, cost--quality trade-offs, and other persistent tensions) \citep{Port96,Levi93,Shak82,McGa23}.

A straightforward numerical example illustrates the challenge of comparing alternatives when performance is multidimensional.  If projects $A$ and $B$ have net present values of, say, \$100 and \$200 million, it is clear that project $B$ is preferable.  That is, finding the best alternative is straightforward: one simply picks the alternative that ranks the highest according to the performance dimension.  But if performance is multidimensional, picking a superior alternative ceases to be straightforward.  Consider the case where project $A$ produces \$100 million and 2 units of social good versus project $B$, which produces \$200 million and 1 unit of social good.  The answer to ``which of the two projects should a firm choose?''\ is no longer self-evident.  Picking among the alternatives now requires comparing two vectors---$(\$100, 2)$ and $(\$200, 1)$---and vectors, unlike one-dimensional quantities, do not have a well-defined ``greater than'' operator, a concept mathematicians refer to as the absence of a ``total order'' (\citealt{Halm74}, Section~14).  In other words, how to compare different alternatives whose performance is multidimensional is not generally defined.

This is problematic, as comparing and choosing among alternatives whose performance is multidimensional is at the core of strategic decision-making.  For instance, the search literature suggests that firms compare nearby alternatives to the status quo and adopt a new strategy if it moves the firm ``up'' along the performance dimension of interest (\citealt{Levi97}; for a survey, see \citealt{Baum19}).  However, as seen with the previous numerical example, there is no ``up'' or ``down'' when comparing multidimensional performance outcomes.  This is not exclusive to the behavioral understanding of firms as searching on a landscape; it also applies to the economic understanding of firms as optimizers, as optimizing also relies on comparing alternatives.

Because of this, economists have advised against even trying to optimize along multiple objectives.  \citet[\p 238]{Jens02} argues that ``[i]t is logically impossible to maximize in more than one dimension at the same time unless the dimensions are monotone transformations of one another'' as an additional reason why ``shareholder primacy'' \citep{Frie70}---just focusing on improving financial performance---is the appropriate course of action for firms.

The CSR literature provides a compelling, real-world illustration of this multidimensionality through its focus on two key performance metrics: corporate financial performance (CFP) and corporate social performance (CSP).\footnote{We acknowledge that treating CSP as a single dimension is a simplification.  A true multi-stakeholder perspective would imply multiple social dimensions---perhaps one for each stakeholder group.  However, we adopt this two-dimensional framing for tractability while preserving generality.  Moreover, in many CSR settings, social performance can be relatively well-defined (e.g., CO$_2$ emissions for airlines), making this simplification reasonable for our purposes.}  Numerous studies have examined whether social and financial performance are positively related by measuring the correlation between these metrics.  This extensive body of research has been summarized by several meta-analyses \citep{Marg03,Orli03,Allo05,Wang16b,Gall19,Vish20}.  The main finding is that, on average, there is a small positive correlation between social and financial performance.  The most recent and comprehensive meta-analysis \citep{Vish20}, covering 344 individual studies, finds a mean correlation of 0.07 (with the 95\% confidence interval going from $-0.14$ to 0.29) and reveals substantial variation across industries (e.g., a correlation of 0.08 in manufacturing and 0.23 in financial services).

The correlation between CFP and CSP is shaped by factors---including technology, consumer preferences, regulations, and stakeholder dynamics---that vary across industries.  In some settings, that correlation is negative, reflecting a trade-off between financial and social goals.  For example, \citet{Hart97b} argue that when ``quality'' (aligned with social outcomes) is hard to contract for, firms face a cost--quality trade-off: cutting costs to boost financial performance can undermine quality, leading to negative CFP--CSP correlations, especially in sectors where safety, environmental protection, or inclusiveness are costly and hard to monitor.

Conversely, CFP and CSP may be positively correlated when firms identify win-win opportunities that create shared value.  \citet{Flam15} finds support for a positive relationship between CSR policies and financial performance.  In her study of large companies that voted on CSR policy-related shareholder proposals and narrowly adopted them, these companies achieved greater financial performance afterward compared to firms that narrowly rejected these proposals.  Similarly, \citet{Doro17} also support a positive relationship between stakeholder reactions and financial performance.  In their study of critical events in small gold mines, those associated with more positive stakeholder reactions in response to critical events showed greater financial performance, while more negative stakeholder reactions led to lower performance.

In other sectors, the CFP--CSP correlation is weak because the activities driving social and financial performance may operate largely independently.  For example, a bank's financial performance depends primarily on its core lending and investment activities, while its social performance often stems from separate philanthropic initiatives with minimal interaction between the two domains.  Additional factors contributing to low CFP--CSP correlation include firms facing an evolving array of stakeholder demands \citep{McGa23} and significant variability in how organizations measure CSP \citep{Kroe14,Rawh17}.

\subsection{Multi-Objective Decision-Making}

While the traditional understanding of maximization does not directly apply to multiple objectives, the literature on multi-objective optimization offers a formal framework for conceptualizing what it means to ``maximize'' performance across multiple dimensions \citep{Keen76,Ehrg05,Emme18}.  This framework provides important distinctions that allow for a more precise and formal language to describe and analyze multidimensional performance.  By introducing concepts such as domination and the efficient frontier, it gives us the tools needed to rigorously compare alternatives across multiple dimensions.  These distinctions serve as the analytical foundation for the search heuristics and modeling approach developed in the remainder of the paper.

\begin{figure}\centering
\includegraphics[width=0.5625\linewidth]{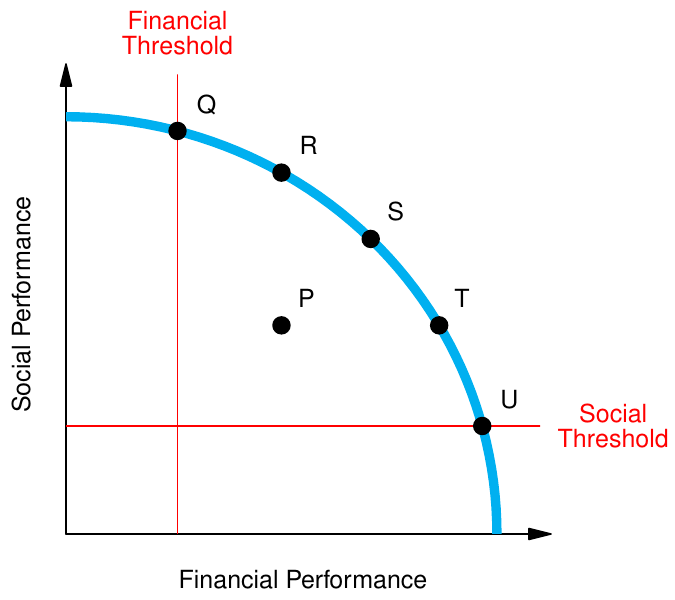}
\caption{The social--financial frontier.}
\label{fig:sf-frontier}
\end{figure}

A foundational idea in this literature is representing performance as a point in a multidimensional space.  For CSR, with its dual focus on social and financial outcomes, a firm's strategy corresponds to a point on a two-dimensional plane.  The set of all feasible strategies forms a region in this space, as depicted in Figure~\ref{fig:sf-frontier}.\footnote{If the space of strategies is not continuous, instead of a ``region'' it would be a ``discrete set of points.''  We use the former term for succinctness.}

In the plane of possible strategies, some positions are objectively inferior to others, as they perform worse in both dimensions (a relationship called \emph{strict domination}) or perform worse in one dimension while being at parity in the other (\emph{weak domination}).  For instance, in Figure~\ref{fig:sf-frontier}, $S$ strictly dominates $P$, while both $R$ and $T$ weakly dominate $P$.  Some points are not dominated by any other points (in the example, all points except $P$ are not dominated by any other point).  The set of all non-dominated points is called the \emph{efficient frontier} \citep[\p 70]{Keen76}.

Because firms often lack complete knowledge of the frontier, they strive for an ``ideal point'' reflecting their preferences for social and financial outcomes.  Due to technological and practical limitations, the best achievable outcome is the ``best viable point''---the point on the frontier closest to the ideal.  Firms' locations on the frontier can represent a range of strategic positions.  For instance, a solely financially focused firm might aim for point $U$, a socially focused foundation for point $Q$, and firms balancing diverse stakeholders somewhere in between, like point $S$.  Firms are limited in where they locate by financial and social thresholds.  Below the financial threshold, a firm will go bankrupt, and below the social threshold, the firm is not meeting its social goals.

\begin{figure}\centering
\includegraphics[width=\linewidth]{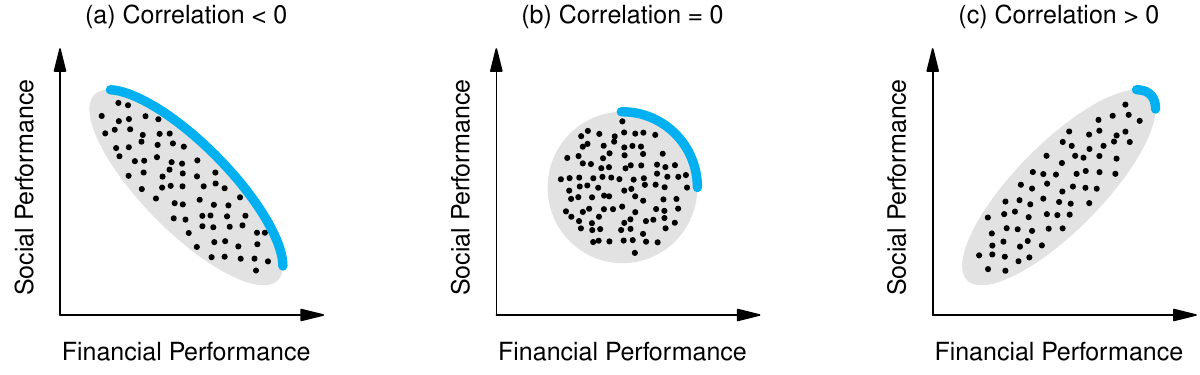}
\caption{The effect of correlation on the efficient frontier.  Panels (a)--(c) show negative, zero, and positive correlations, respectively, between firms' social and financial performance; black dots represent individual firms.  The arcs represent stylized efficient frontiers.}
\label{fig:frontiers}
\end{figure}

The correlation between performance dimensions, which meta-analyses have shown to vary substantially across industries (recall the 95\% confidence interval of correlation going from $-0.14$ to 0.29), significantly influences the shape and size of the feasible region, and consequently, the efficient frontier \citep{Adne14}; see Figure~\ref{fig:frontiers} for an illustration.  For instance, in industries where the meta-analyses show negative correlations between CSP and CFP, such as certain manufacturing sectors, reducing environmental impact may often come at a financial cost, resulting in a backward-sloping set of positions and a larger frontier (see panel~(a)).  Conversely, in industries like financial services where meta-analyses reveal stronger positive correlations, improvements in social and financial performance tend to be mutually reinforcing, leading to a relatively smaller frontier (see panel~(c)).  Industries with correlations close to zero, which align with the overall mean correlation of 0.07 found in the literature, might allow for a wider range of social--financial combinations (see panel~(b)).

Correlation and trade-offs are inversely related: as correlation increases, the prevalence of trade-offs diminishes.  Trade-offs arise when a firm cannot improve along one dimension without sacrificing performance along another.  This situation occurs for firms that are \emph{on the frontier} (the highlighted arcs in Figure~\ref{fig:frontiers}).  As correlation increases---that is, as one moves from panel (a) to panel (c)---the size of the frontier shrinks.  Consequently, the number of firms facing trade-offs (those on the frontier) decreases as correlation increases.

\subsection{Multi-Objective Search}

The previous subsection established how to conceptualize ``optimizing'' with multiple objectives for a given set of alternatives.  However, it does not address the \emph{process} by which firms might discover high-performing (or frontier) points in the first place.  The behavioral theory of the firm, grounded in the concept of bounded rationality \citep{Simo55}, is critical here.  Firms face significant cognitive and informational limitations; they cannot possibly know all potential strategies and their associated outcomes across all relevant performance dimensions.  Instead of comprehensively evaluating all possibilities, firms engage in \emph{search} \citep{Cyer63,Levi97}: they must generate and discover alternative strategies before they can compare these to the status quo.  Decision-making, by contrast, is the selection among currently available alternatives.  Search is thus logically prior to decision-making, and it is path dependent: once a new strategy is adopted, it becomes the status quo that shapes which neighboring alternatives the organization is likely to discover next.

This distinction matters especially under multiple goals.  With multiple performance dimensions and potential trade-offs, preferences do not only determine which of the \emph{known} alternatives a decision maker selects; they also shape how search proceeds.  Put differently, preferences must be \emph{implemented} in some procedure for evaluating candidate moves during search, and that procedure influences which regions of the performance space the firm ever reaches.  In practice, search and choice are not neatly sequential: each step of exploration creates familiarity and commitment that shape subsequent evaluation \citep{Gans19}, so the evaluation rule is not applied once to a fixed menu but repeatedly, steering the trajectory of discovery itself.

\emph{The classical NK landscape framework}.  The NK landscape framework \citep{Levi97} offers a powerful way to model how firms evaluate and ultimately search for better strategies.  Each point on the landscape represents a specific combination of decisions, with its ``height'' indicating performance.

To illustrate, consider a car manufacturer facing three key decisions: (i) engine type (electric or combustion), (ii) car type (SUV or pickup), and (iii) market (US or global).  Each of the eight possible combinations of these choices yields a different outcome in terms of performance.  For example, selling combustion SUVs globally might maximize profits, and selling electric pickups in the US might be less profitable.

The parameter $N$ in the NK model represents the number of such strategic decisions the firm makes.  The parameter $K$ captures how much these decisions interact: for instance, the profitability of selling SUVs may depend on whether the target market is the US or global, but not on whether the cars are electric or not.  When $K$ is high, these decisions are highly interdependent---changing one aspect (e.g., switching to electric engines) can have ripple effects on performance depending on the other choices made.  This interconnectedness creates a ``rugged landscape,'' where small changes can lead to unpredictable jumps in performance.

Due to bounded rationality, managers cannot foresee the consequences of all possible combinations but can experiment with small changes to observe their effects.  Thus, firms explore a rugged landscape by using local search---making incremental changes to their decisions and moving to a neighboring strategy only if it improves performance according to their chosen heuristic.

\begin{figure}\centering
\includegraphics[width=0.7\linewidth]{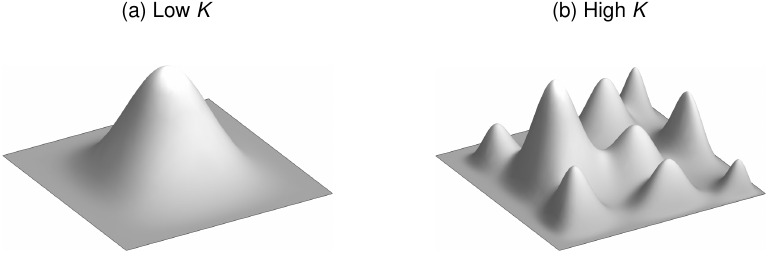}
\caption{Illustration of landscape complexity ($K$) in a unidimensional setting.  Higher $K$ leads to a more rugged landscape with multiple local peaks.}
\label{fig:landscapes}
\end{figure}

The effect of $K$ can be visualized as increasing the ruggedness of a three-dimensional landscape (see Figure~\ref{fig:landscapes}).  Although this topographical metaphor is not fully accurate, it effectively illustrates a key insight from this literature: as complexity grows, boundedly rational actors---those who can only search their immediate neighborhood---find it increasingly difficult to reach the global peak.  For readers interested in a deeper understanding of NK models, we recommend \citet{Csas18c}, which provides an accessible introduction to this rich literature.

A key result of the literature on rugged landscapes is that firm performance depends on the interaction between the search heuristic used by the firm and the characteristics of the landscape.  This interplay embodies Simon's \citeyearpar{Simo90b} ``scissors'':\ the search outcome depends on the interaction between the task environment and the decision maker.

However, the role that search plays under multiple goals has received limited attention in the search literature, and the NK literature in particular, despite strong reasons to believe it matters.\footnote{An important exception is \citet{Ethi04}, who model search under multiple goals as a constraint: any change must not decrease performance on other goals.  Their work demonstrates that strict improvement requirements can limit adaptability, but focuses less on how different approaches to evaluating multiple goals shape search trajectories.}  This is surprising, given that the management of multiple goals in organizations constitutes a central pillar of the behavioral theory of the firm \citep{Cyer63}.

Empirical research suggests that firms differ in how they prioritize and manage such trade-offs---and that these differences have significant consequences for effective search.  For example, \citet{Grev08} shows that insurance firms balance size and profitability goals, with firms below their size targets growing faster---especially when profitability goals are met.  \citet{Gaba19} find that airlines prioritize safety and profitability based on survival needs: firms with low profitability focus more on safety, while highly profitable firms are less responsive to safety concerns.  \citet{Albe26} show that organizational subgroups representing distinct goals---revenues versus patient care in hospitals---may pursue the same objectives differently depending on their hierarchical position.  Given such variation in how organizations manage multiple goals, understanding how they search under multiple goals---and whether different approaches lead to different outcomes---becomes an important question.

\emph{Extending to two performance landscapes (dual landscape model)}.  To address the single-goal limitation of the NK literature, we build on the dual landscape framework, which allows each strategy to map to multiple performance values.

Returning to our car manufacturer example: each combination of decisions (engine type, car type, market) not only determines a single performance value, but actually maps to \emph{two} performance values: one financial (e.g., profit) and one social (e.g., environmental impact).  For instance, producing electric SUVs for the global market may yield high social performance (due to low emissions) but moderate financial performance if costs are high or consumer demand is uncertain.  Conversely, combustion pickups in the US market may yield high financial returns but poor social outcomes.

\begin{figure}\centering
\includegraphics[width=\linewidth]{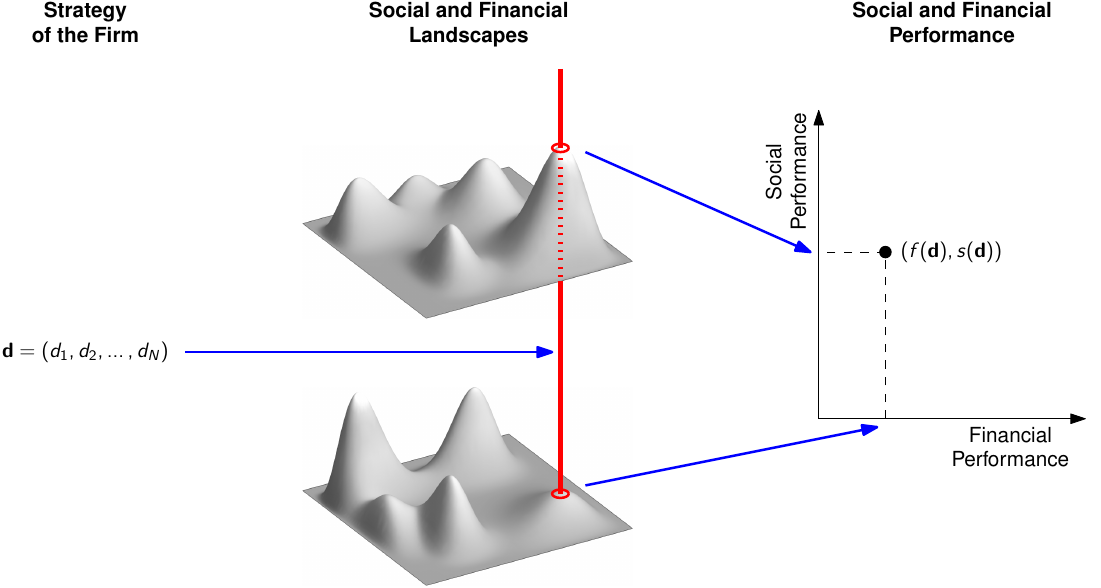}
\caption{A dual-performance landscape.  Each strategy (represented by the vector of decisions $\vec{d}$) has a corresponding performance on the financial and social performance landscapes.  The point $(f(\vec{d}),s(\vec{d}))$ represents the multidimensional performance of strategy $\vec{d}$.  The correlation between landscapes varies by industry and context.}
\label{fig:dual-landscape}
\end{figure}

This dual mapping is captured in the \emph{dual landscape} framework introduced by \citet{Adne14}, in which each strategy is associated with two separate performance values---one on a financial landscape and one on a social landscape (see Figure~\ref{fig:dual-landscape}).  While \citet{Adne14} originally developed the dual landscape framework to examine industrial organization questions, it provides an ideal foundation for our analysis of how organizations search under multiple objectives.  The degree to which these two landscapes (or objectives) are aligned or in conflict is governed by the correlation parameter ($\rho$): in some industries, strategies that increase financial performance also tend to improve social outcomes (positive correlation), while in others, the opposite may happen (negative correlation).

Figure~\ref{fig:dual-landscape} illustrates how a single strategy---a particular combination of decisions---maps to both financial and social performance.  Importantly, there is only \emph{one} underlying set of decisions: the same interdependent decision vector $\vec{d}$ jointly determines both outcomes, so changing any component of $\vec{d}$ can have coupled consequences for financial and social performance.  When the two landscapes are highly correlated, strategies that do well financially also tend to do well socially.  When the correlation is negative, improving one dimension often comes at the expense of the other, making it difficult to find strategies that perform well on both.

\subsection{Search Heuristics for Social and Financial Performance}

To understand search in a multidimensional setting, we must consider how firms generate and select alternatives.  The behavioral theory of the firm suggests that search is often local and iterative, informed by a firm's existing knowledge \citep[e.g.,][]{Kati02,Rose01,Flem01}, primarily promoting incremental change that is implemented if decision makers consider it superior to the status quo.  That is, firms generally search for improvements to their current strategy rather than for entirely different ones.  When evaluating such alternatives in a one-dimensional performance context, the heuristic applied is one of ``greater than'' (up or down).  However, because ``up'' cannot be unambiguously defined in a multidimensional setting, other heuristics must take the place of the unidimensional comparison operator.

Given our focus on how organizations balance multiple stakeholder interests, and recognizing that CSR decisions (involving both social and financial performance) represent a canonical example of this challenge, we draw on the CSR and multiple-goals literatures to identify relevant heuristics.

To understand search in a multidimensional setting, it is important to distinguish between preferences and heuristics.  \emph{Preferences} represent a firm's desired position in the performance space---for instance, aiming for point $S$ in Figure~\ref{fig:sf-frontier} would reflect equal weighting of social and financial outcomes.  However, preferences alone do not determine how firms move toward their goals.  That process is governed by \emph{heuristics}---the specific rules and procedures firms use to evaluate strategies that align with their preferences.  Multiple heuristics can serve the same preference structure.  For example, a firm equally valuing social and financial performance could employ either a Combine heuristic that evaluates both dimensions simultaneously or an Alternate heuristic that switches focus between them, as both heuristics attend to both dimensions.\footnote{The selection between these heuristics may also be influenced by other factors, such as a firm's preference for stable choices or the cost of implementing a given heuristic.}  This distinction matters because, as we will demonstrate, different heuristics vary significantly in their effectiveness at achieving the same preferences: the same ``ideal point'' can be pursued through different search procedures that generate different search trajectories and, consequently, different outcomes.

Drawing from the CSR and multiple-goals literatures discussed above, we identify five distinct heuristics that capture the main ways organizations approach the dual pursuit of social and financial performance.  Two of these are \emph{compensatory} in the sense that they explicitly trade off the two dimensions on a single evaluative scale (Combine and Penalize), which typically requires commensurability of social and financial performance \citep{Carr99,Espe98}.\footnote{We thank an anonymous reviewer for encouraging us to think about the role and implications of commensurability in our model.}  Two are \emph{non-compensatory} in that they avoid aggregation by focusing attention on one dimension at a time (Alternate and Satisfice), thereby reducing commensurability demands to within-dimension comparisons and/or threshold assessments.  Finally, one represents a profit-first baseline that disregards social performance (Maximize).  To clarify the logic of each heuristic, we first describe the rationale behind each approach---the core decision-making principle guiding search---before formally modeling them in the next section.

\emph{1.~Maximize financial performance.}  Under this rationale, decision-makers ignore social performance and focus entirely on trying to maximize financial performance.  Consequently, this rationale reflects a preference structure that puts zero weight on social performance and all the weight on financial performance.  This view is closest to Friedman's \citeyearpar{Frie70} doctrine that ``the social responsibility of business is to increase its profits.''  Although few firms may claim to completely disregard social performance outcomes, the profit-maximizing perspective has historically been a ``dominant logic'' \citep{Prah86,Cril12} shaping corporate behavior, as evidenced by the Business Roundtable's only recent (2019) shift toward acknowledging commitments to all stakeholders \citep{Beno19}.  We argue that this heuristic poses a meaningful benchmark that allows us to evaluate the opportunity costs or gains associated with using any of the multi-objective heuristics in the CSR context.  For instance, some hedge funds may evaluate investments solely based on risk-adjusted returns, where social considerations matter only if they affect those returns.

\emph{2.~Combine social and financial performance.}  Under this rationale, decision-makers evaluate social and financial performance simultaneously by using a composite measure that merges both dimensions into a single metric.  This approach reflects a decision maker who assigns specific weights to each type of performance, rather than prioritizing one exclusively.  Behavioral theory refers to such decision processes as ``additive,'' since they sum performance across multiple dimensions (\citealt[\p 341]{Brom14}; \citealt[\p 651]{Gaba19}).  Case studies in the CSR literature show that organizations implement this rationale by encouraging employees to seek compromises among different objectives \citep[\p 129]{Batt19}.  Because this approach aggregates social and financial performance, it bears similarities to utilitarian ethics, which seek to maximize total welfare across all relevant individuals \citep{Bent1789,Mill1848}.  A possible example is a social enterprise operating under a ``buy-one-give-one'' model, where business decisions are evaluated through both commercial and social consequence criteria simultaneously \citep[e.g.,][]{Marq14}.

Notably, combining social and financial performance in this manner requires that these dimensions be commensurable---that is, capable of being measured and aggregated on a common scale \citep{Carr99,Espe98}.  The logic of this heuristic relies on the ability to merge distinct domains into a unified metric, a process that mirrors real-world practices such as ESG (Environmental, Social, and Governance) scoring systems and triple bottom line accounting.  Both of these approaches aggregate multiple performance dimensions into a single evaluative score, using explicit or implicit weighting schemes.  However, achieving such commensurability is challenging.  The literature on social impact measurement highlights the lack of consensus on how to measure or compare social value creation, which is itself a multidimensional and complex construct \citep{Kroe14,Rawh17}.  As a result, firms may adopt different methods for combining social and financial performance.

\emph{3.~Alternate between pursuing social and financial performance.}  This rationale embodies the observation in the behavioral theory of the firm that organizations often focus on one goal at a time \citep[\p 118]{Cyer63}.  Like Combine, it can reflect a firm's joint preference for both dimensions.  From time to time, organizations switch from one goal to another, typically when progress along one goal has plateaued and that goal therefore ceases to be salient (\citealt[\p 480]{Grev08}, \citealt[\p 344]{Brom14}).  In the context of CSR, \citet{McGu88} provide some evidence consistent with this rationale by finding that social investments are often preceded by significant improvements in financial performance.  In a similar vein, \citet[\p 315]{Tant16} point out that managers often deal with CSR pressures by rotating their attention among different trade-offs.  For example, a hospital may, for an extended period, prioritize financial viability (profitability) as rising costs place increasing pressure on operations.  Over time, however, management may shift its focus toward population health outcomes, that is, a mission-driven objective that includes, for instance, expanding charity care for underserved patients---a shift that could be triggered by heightened social pressures, leadership turnover, or changes in governance (e.g., takeover by a private equity owner).  This ``alternating'' approach is closely related to the idea of vacillation in organizational choices, where firms dynamically shift their focus between alternatives over time \citep{Nick02}.  Furthermore, the alternation heuristic shares conceptual similarities with ``mixed strategies'' in non-cooperative game theory, where agents randomize their actions across alternatives to achieve desired outcomes \citep{Ghem85,Shin13}.  However, while mixed strategies are typically stochastic and formalized in game-theoretic terms, the alternating heuristic described here is deterministic and rooted in behavioral theories of sequential attention to goals.\footnote{Please note that the reported results are still stochastic for Alternate in the sense of Monte Carlo simulations.}

\emph{4.~Penalize social underperformance.}  This rationale builds on the logic outlined by \citet{McWi01}.  According to this approach, firms assess different alternatives by considering a modified measure of financial performance that takes into account their desired level of social performance.  For instance, a car manufacturer may strive to produce cars with zero carbon footprint.  When evaluating new car designs, the company will penalize designs that fall short of this desired carbon footprint.  This rationale reflects the decision-makers' aim to maximize financial performance while also ensuring a desired level of social performance.  Note that penalizing social underperformance differs from maximizing financial performance because it not only accounts for financial performance (which already incorporates the costs or benefits associated with the firm's social performance) but it also takes into account the failure to meet a self-imposed social threshold.  For example, a manufacturer may conclude that because its social performance goals were not met, overall performance should be discounted---even if financial and operational targets were nominally achieved.  This ``recalibration'' can, for instance, reduce bonus payments for decision makers.  Like the Combine heuristic, Penalize requires that social and financial dimensions be made commensurable so that shortfalls in social performance can be translated into financial penalties and aggregated into a single evaluative metric \citep{Carr99,Espe98}.  Because of this, this approach is also consistent with utilitarianism \citep{Bent1789,Mill1848}.

\emph{5.~Satisfice social performance.}  Under this rationale, firms prioritize financial performance as long as an aspired social threshold is cleared.  Similar to Penalize, this logic reflects a decision maker's preference for financial performance, conditional upon achieving their chosen level of social performance.  When the social threshold is not cleared, the firm focuses solely on improving social performance.  This rationale has its roots in the behavioral theory of the firm; in particular, the argument that firms engage in problemistic search when they underperform their aspiration on a specific criterion (\citealt[\pp 161--162]{Marc58}, \citealt[\pp 119--120]{Cyer63}).  In the context of CSR activities, this implies that attempts to enhance social performance cease once the firm has attained a satisfactory social outcome.  The focus then shifts toward enhancing financial performance.

This logic is reflected in practices such as B Corp certification, regulatory compliance, or industry codes of conduct, where firms must meet minimum standards (e.g., for labor rights, environmental impact, or diversity)---or aim to achieve certain certifications (e.g., ISO standards)---before pursuing additional financial or strategic objectives.  Importantly, this approach is closely related to the concept of lexicographic preferences in economics, where decision-makers prioritize one criterion absolutely over others, only considering secondary dimensions once the more relevant dimensions have been satisfied \citep{Enca64,Mold92}.  In this way, the satisficing heuristic reflects a lexicographic ordering of objectives, in contrast to the compensatory logic of utilitarian or additive approaches.  Unlike Combine and Penalize, Satisfice does not require translating social performance into financial units; rather, it requires that the decision maker can (i) assess whether the social threshold is met and (ii) evaluate improvements within the currently focal dimension.  For example, a firm that intends to satisfice on labor practices may easily assess whether the minimum standards are met (e.g., a cap on working hours per day) without measuring worker welfare in financially commensurable terms or measuring welfare beyond that threshold.

These five heuristics represent distinct approaches to navigating the multidimensional performance challenges prevalent in the CSR context.  In the following sections, we formally model these heuristics, explore their performance under varying conditions, and offer insights into their relative effectiveness and the trade-offs they entail.

The focus of the analyses in the ensuing sections is to understand the consequences of using these different heuristics (i.e., their performance outcomes) rather than their antecedents (i.e., why organizations adopt one heuristic over another).  This latter question, while important, falls outside the scope of our paper.  However, it is worth noting that heuristic adoption is likely influenced by several factors, including incentive conflicts between principals and agents \citep{Jens76} and the tendency to focus on outcomes that are more easily measured and rewarded \citep{Kerr75,Ethi09}.  For instance, managers may gravitate toward heuristics that emphasize financial performance when social outcomes are difficult to measure or when compensation structures primarily reward financial results.  These complexities in organizational decision-making further underscore the importance of understanding how different heuristics perform once adopted.

\section{Model}\label{sec:model}

To understand the effect of search heuristics on firms' social and financial performance, we develop a model to compare the performance of the five search heuristics described above when applied to a multi-attribute landscape \citep{Adne14}.  The remainder of this section (i) formally describes the five search heuristics and (ii) explains how we simulate and compare their social and financial performance.

\subsection{Modeling Dual-Goal Search Heuristics}

The heuristics we model capture the five decision-making rationales mentioned in the previous section: \emph{Maximize} financial performance, \emph{Combine} social and financial performance, \emph{Alternate} between pursuing social and financial performance, \emph{Penalize} social underperformance, and \emph{Satisfice} social performance.  From here on, we simply call these heuristics Maximize, Combine, Alternate, Penalize, and Satisfice.  Our heuristics can further be categorized by the preferences they inherently reflect, that is, single dimension preference (Maximize), joint preference for social and financial performance (Combine and Alternate), and contingent preferences (Penalize and Satisfice).

Following prior research, we model firms as boundedly rational agents \citep{Simo55}, whose search for better strategies is inherently local and incremental, centered around modifications to their current strategy \citep{Cyer63,Levi97}.  At each time period, each firm in our model performs local search, comparing its performance to the performance of the neighboring strategies.  A period represents one evaluation/choice cycle (e.g., a stage-gate, KPI review, budgeting round, or managerial attention cycle), not a calendar year.  For simplicity, we assume that strategy vectors only include 0s and 1s; thus, there are $N$ neighboring strategies (e.g., if the current strategy is $(0,0,0)$, there are three neighboring strategies: $(1,0,0)$, $(0,1,0)$, and $(0,0,1)$).

\begin{figure}\centering
\includegraphics[width=\linewidth]{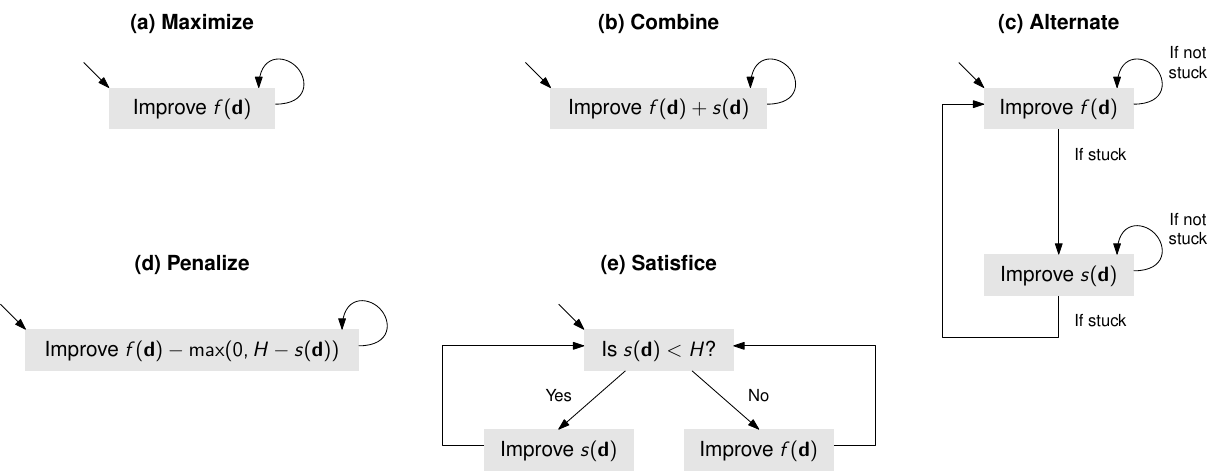}
\caption{Flow diagrams describing the five search heuristics we analyze.}\label{fig:heuristics}
\end{figure}

How performances are compared depends on the heuristic.  Each strategy $\vec{d}$ has a financial value $f(\vec{d})$ and a social value $s(\vec{d})$, which we normalize to $[0,1]$ for comparability across runs.  Figure~\ref{fig:heuristics} summarizes the five heuristics we analyze using the following notation: $\vec{d}$ is the firm's decision vector; $f(\vec{d})$ and $s(\vec{d})$ are its financial and social performance (the ``height'' on each landscape in Figure~\ref{fig:dual-landscape}); and $H$ is the social threshold.  The ``improve'' boxes denote one step of local search: the firm evaluates its $N$ one-bit neighbors and moves if the heuristic identifies a superior neighbor.  Importantly, only the compensatory heuristics (Combine and Penalize) require cross-dimensional commensuration---i.e., an ability to aggregate or trade off $f(\vec{d})$ and $s(\vec{d})$ in a single evaluative criterion---whereas Alternate and Satisfice rely on within-dimension comparisons and/or a threshold check, imposing weaker commensurability demands (we return to these implementation implications in the Discussion).

\emph{Maximize} (panel~(a)).  This heuristic only measures performance in terms of $f(\vec{d})$ and, thus, is indifferent to social performance.  At each time period, Maximize tries to improve financial performance by performing a local search on the financial landscape.  Maximize provides a baseline against which to compare the remaining heuristics.

\emph{Combine} (panel~(b)).  This heuristic measures performance as the sum of financial and social performance.  That is, at each time period, a firm using this heuristic tests all neighboring strategies and picks the one that has the highest sum of social and financial performance (that is, it makes choices based on $f(\vec{d}) + s(\vec{d})$).\footnote{This rule is different from Ethiraj and Levinthal's \citeyearpar[\p 14]{Ethi09} multi-objective benchmark rule, which only moves to a new position if that position weakly dominates the current position (that is, if no performance dimension declines).  A more fundamental difference with \citet{Ethi09} is that they only examine the effect of search heuristics on one performance measure (fitness).}  In Appendix~\ref{app:weighted-combine}, we explore the more general class of ``Weighted Combine'' heuristics, where performance is measured as the weighted sum $w f(\vec{d}) + (1-w) s(\vec{d})$, with $w$ in $[0,1]$.

\emph{Alternate} (panel~(c)).  This heuristic works on one goal at a time.  It picks a first goal at random and then continues trying to improve that goal for as long as this is possible (that is, until ``stuck'' at a local peak).  Once improvement on a given goal ceases to be possible, the heuristic switches to trying to improve the other goal, and so on.

\emph{Penalize} (panel~(d)).  This heuristic measures performance as financial performance minus a penalty if the firm is underperforming the social threshold $H$.  The penalty is the amount of social underperformance, $\max(0, H - s(\vec{d}))$.

\emph{Satisfice} (panel~(e)).  This heuristic tries to improve social performance as long as that performance is below the social threshold $H$; otherwise, it attempts to improve financial performance.

Note that it is possible to model other heuristics apart from the five just described.  For instance, one could model the mirror images of Maximize and Satisfice (i.e., ``maximize social performance'' and ``satisfice financial performance'').  We focus on the five heuristics described so far to keep the presentation of the results tractable and because they best illustrate the five logics highlighted by the CSR and multiple goals literatures.  In any case, once one understands the behavior of these five heuristics, it is not difficult to predict the behavior of other heuristics.

\subsection{Measuring Performance via Simulation}

To compare the performance of the five search heuristics, we proceed as follows.  First, we create a dual-performance landscape with complexity $K$ and correlation $\rho$ (an overview of this method was presented in the Theoretical Background section; for details see \citealt[\pp 2796--2798]{Adne14}).  To make results comparable across simulations, we scale social and financial performance so that 0 corresponds to the global minimum and 1 to the global maximum of its respective landscape.  This makes it easier to interpret our comparisons.

Second, we create five firms that start at the same random point on the landscape.  Each firm uses exclusively one of the five heuristics for its search over the multi-attribute landscape.  Third, we let the firms search for 200 time periods.  We chose this number because, by then, each firm's average performance has plateaued.  We record each firm's social and financial performance in period 200.  Finally, we repeat the previous three steps 10,000 times for each combination of parameters we explore.  This allows us to estimate the expected social and financial performance of each heuristic given complexity ($K$), correlation ($\rho$), and social threshold ($H$).  In additional analyses reported in the Appendix~\ref{app:random} (and summarized in the Results section), we partially relax strict local search by allowing occasional non-local (random) moves during search.

\section{Results}\label{sec:results}

We report our results using scatterplots that allow for an intuitive and precise description of the model's behavior.  The points in these scatterplots show the different heuristics measured against social and financial performance (on the vertical and horizontal axes, respectively).  Unless otherwise noted, each point represents average performance at the end of the simulation ($T=200$) over 10,000 Monte Carlo trials for a given parameter setting.

The particular values used to generate the plots were chosen because they illustrate the whole behavior of the model.  We keep the number of decisions fixed at $N=12$, as different values do not change the behavior of our model qualitatively as long as complexity ($K$) is scaled with $N$.  For complexity $K$, we plot low, medium, and high values ($K=1$, 6, and 11) and for correlation $\rho$, we plot negative, neutral, and positive values ($\rho=-0.5$, 0, and 0.5).  For the social threshold $H$, our first analyses use a medium value ($H=0.5$), which we later expand to include low and high values ($H=0.2$ and 0.8).  Understanding our model's behavior under this set of values is enough to understand its whole behavior.

We structure the exposition of the results as follows: We first describe general patterns that are robust to parameter changes.  We then explain what drives the expected performance of each heuristic and we examine the variability of the results.  Finally, we examine the conditions under which some heuristics can match or outperform the social and financial performance of Maximize.

\subsection{General Patterns: The Effect of Complexity and Correlation}

\begin{figure}\centering
\includegraphics[width=\linewidth]{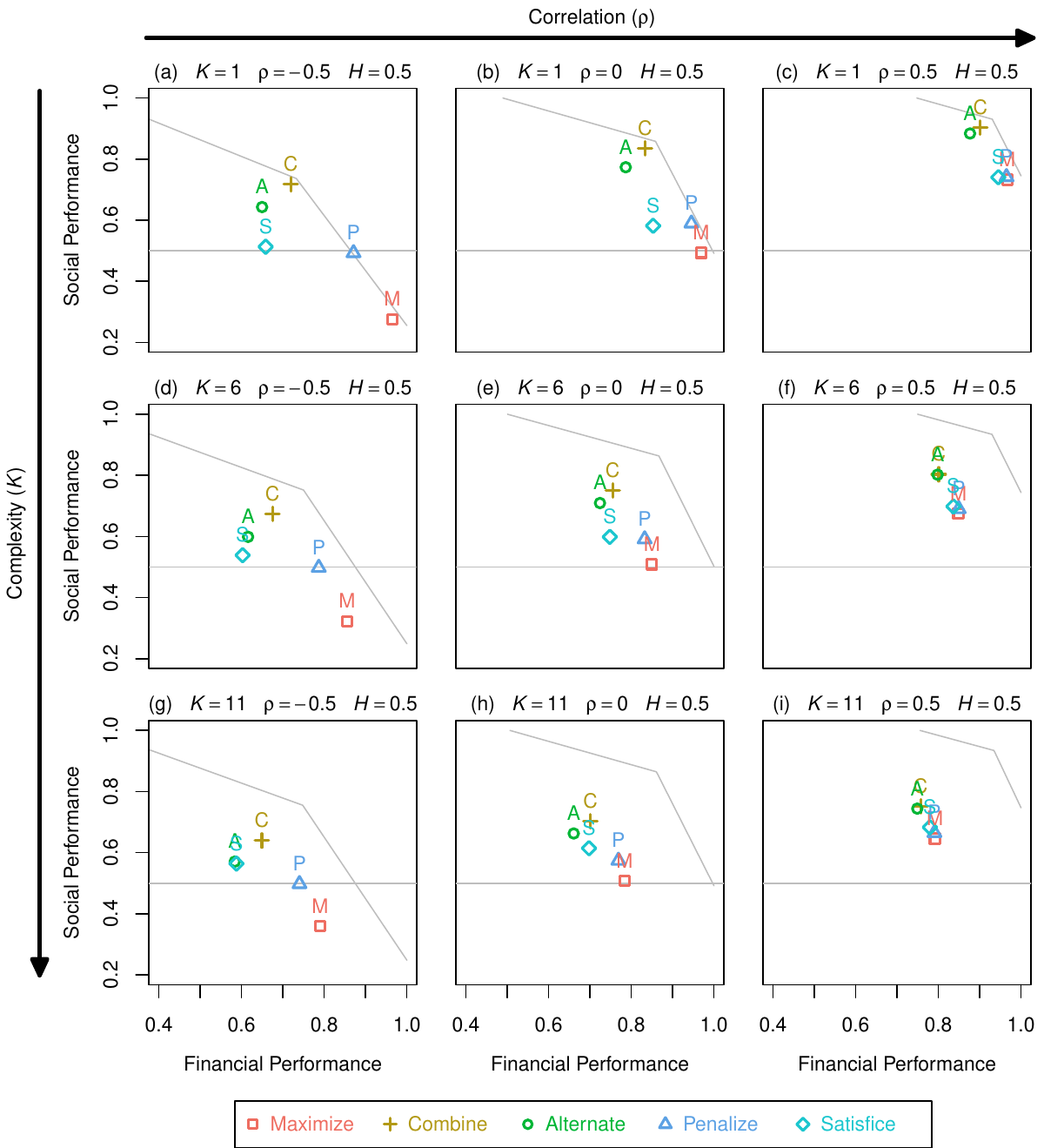}
\caption{Performance plots for the heuristics as a function of complexity ($K$) and correlation ($\rho$).}\label{fig:performance-k-rho}
\end{figure}

Figure~\ref{fig:performance-k-rho} shows nine performance plots (labeled (a) through (i)), which differ in complexity (row-wise) and correlation (column-wise).  For all nine plots, the social performance threshold is fixed at $H=0.5$.  Each panel shows the average social and financial performance of each heuristic.  The gray broken line in each panel represents the average position of the efficient frontier in that panel.\footnote{Each frontier is approximated using three points (averaged over 10,000 simulations): maximal financial performance, maximal social performance, and maximal sum of social and financial performance.}

An overarching observation from Figure~\ref{fig:performance-k-rho} is that varying complexity ($K$) and correlation ($\rho$) does not change the relative ordering of the heuristics.\footnote{Here, ``relative ordering'' refers to the ranking of heuristics by their average locations in the social--financial plane and to which heuristics tend to be closer to the top-right, bottom-right, or top-left regions of each panel.}  For example, Maximize is always near the bottom-right corner and Combine is always the closest to the top-right corner.  The reason is that $K$ and $\rho$ affect the landscapes on which search occurs, but they do not alter the directional pull built into each heuristic's acceptance rule.  Maximize follows financial improvements only, leaving social performance to depend on the social value of the financial peak it reaches.  Combine evaluates moves using the joint criterion $f(\vec d)+s(\vec d)$, leading it toward balanced compromise positions.  Alternate switches between financial and social search, moving firms toward locally attractive positions on each landscape.  Penalize and Satisfice both depend on the social threshold $H$, but in different ways: Penalize adjusts the evaluation of financial moves by subtracting social shortfalls, whereas Satisfice switches to social search until the threshold is met.  Thus, $K$ and $\rho$ change how difficult search is, and whether gains in one dimension tend to carry over to the other, but not the basic direction imposed by each heuristic.

This observation suggests that heuristic choice is central to explaining firms' relative positions in the social--financial space.  Within the same environment, different heuristics can lead firms to reach systematically different outcomes, depending on whether they search by maximizing one objective, combining objectives, alternating attention between them, or using a social threshold.  At the same time, environmental structure affects the stakes of heuristic choice.  When $\rho$ is high or $K$ is low, the heuristics cluster more tightly, making the choice among them less consequential within the model.  When $\rho$ is low and $K$ is high, differences across heuristics become larger because joint improvement is harder and local search is more constrained.  Thus, environmental structure shapes the difficulty of joint improvement, whereas heuristic choice shapes where search tends to land within that environment.

As correlation increases, the performance of all heuristics improves along both dimensions.  This is evident in Figure~\ref{fig:performance-k-rho}, where all points move closer to the top-right corner of each panel as one moves from left to right across the columns.  The underlying reason is that higher correlation means that improvement in one dimension is likely to lead to improvement in the other.  For example, if social and financial performance are correlated, a firm that enhances its social performance is also likely to see gains in its financial performance.  In other words, increased correlation strengthens the extent to which progress on one landscape facilitates progress on the other.  This effect is particularly apparent for the Maximize heuristic, which ignores social performance but nevertheless sees improvements in this dimension as correlation rises.  Overall, increasing the correlation between performance dimensions not only boosts performance across all heuristics but also reduces the absolute differences among them.

When complexity increases, the performance of all heuristics decreases along both dimensions.  In Figure~\ref{fig:performance-k-rho}, this can be seen as all points moving farther away from the top-right corner of each panel as one moves row-wise from top to bottom.  The reason is that as complexity increases, the performance landscape on both dimensions becomes more rugged and, hence, harder to search; the landscapes have more local peaks where search can get stuck \citep{Levi97}.  Hence, all heuristics are less likely to find the global peak and instead quickly get stuck near their starting position.

The fact that the ordering of the heuristics is stable across the parameter settings we examine allows us to simplify the exposition of the results and focus on explaining the relative financial and social performance of each heuristic.  Because performance differences are most visible in panel~(a) in Figure~\ref{fig:performance-k-rho}, our descriptions below refer to that panel unless stated otherwise.

\subsection{Relative Performance of the Heuristics That Do Not Depend on a Threshold}

We now move on to describe the relative ordering of the different rules.  For clarity, we first analyze the decision rules that do not take into account the social threshold ($H$); namely, Maximize, Combine, and Alternate.

Maximize achieves the highest financial performance and the lowest social performance across all panels in Figure~\ref{fig:performance-k-rho}.  To understand this behavior, recall that this decision rule performs an unconstrained search over the financial landscape.  Hence, this heuristic devotes all its attention to maximizing financial performance without having to compromise to achieve any particular social performance.  As a result, it is natural that Maximize appears at the bottom-right corner of all panels in Figure~\ref{fig:performance-k-rho}.

The next two heuristics---Combine and Alternate---both reflect a joint preference for social and financial performance, yet they differ with respect to how they pursue these preferences.  While Combine pays attention to both objectives \emph{at the same time}, always looking at the sum of social and financial performance, Alternate pays the same attention to these goals but \emph{over time}, switching its focus back and forth between social and financial performance.  In both cases, joint attention to both performance measures is reflected in the fact that the average performance of both heuristics falls along the 45-degree diagonal of each panel.

Interestingly, Combine comes closer to performing ideally than Alternate; it reaches closer to each panel's top-right corner, where both social and financial performance are maximized.  A priori, there are arguments for either of these to outperform the other.  On the one hand, Combine searches over a landscape that is the sum of the social and the financial landscapes and therefore generally more rugged than either of those.  The higher effective complexity faced by Combine means that it will get stuck quickly; however, it will do so at a peak that represents a compromise between the two performance dimensions.  On the other hand, Alternate will face less rugged landscapes, as it searches one dimension at a time.  However, Alternate is unlikely to settle on any given point: once it has found a local peak on, say, the financial landscape, it will switch to searching on the social landscape, and so on.  That is, Alternate never settles at a point that is a good compromise between the two performance dimensions (unless that point happens to be a local peak in both landscapes).  The result is that Combine finds a good position and stays there, while Alternate continues to explore and the average of that continuous exploration is farther from the ideal than Combine's solution.  We will add nuance to this conclusion later when we discuss the ability of the different heuristics to explore.

\subsection{Relative Performance of the Heuristics That Depend on a Threshold}

\begin{figure}\centering
\includegraphics[width=\linewidth]{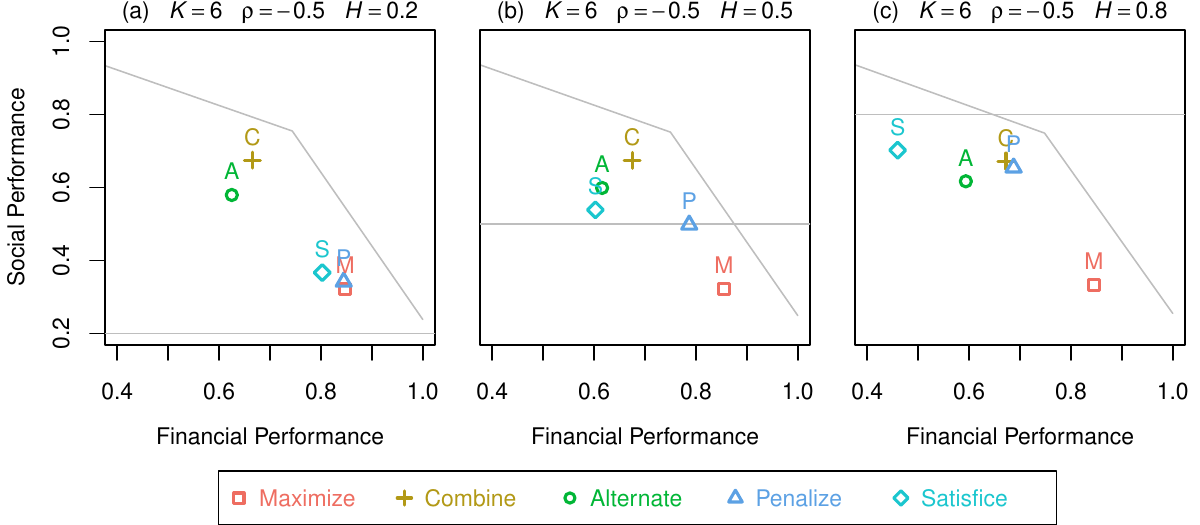}
\caption{Performance plots for different social thresholds ($H$).}\label{fig:performance-h}
\end{figure}

We now analyze the remaining heuristics, Penalize and Satisfice, both of which reflect a preference for financial performance that is contingent on meeting a minimum social performance threshold ($H$).  Here, we will be looking at Figure~\ref{fig:performance-h}, which uses the same $K$ and $\rho$ values as panel~(d) in Figure~\ref{fig:performance-k-rho} while varying the social threshold to take low, medium, and high values ($H=0.2$, 0.5, and 0.8).

The Penalize heuristic moves from being close to Maximize to being close to Combine as the social threshold increases (i.e., as one moves from panel~(a) to panel~(c) in Figure~\ref{fig:performance-h}).  This happens because the threshold affects the topography of the penalty-adjusted landscape that this heuristic searches.  When $H=0$, there is never a penalty associated with social performance, as it is impossible to underperform this threshold.  Hence, the penalty-adjusted landscape is simply the financial landscape and, thus, the Penalize heuristic behaves identically to Maximize (that is, if $H=0$, then $f(\vec{d}) - \max(0, H-s(\vec{d}))$ reduces to $f(\vec{d})$).  Conversely, when $H=1$, social performance is always underperforming the threshold, making the penalty-adjusted landscape an even mix of social and financial performance.  Penalize then behaves identically to Combine (that is, if $H=1$, then $f(\vec{d}) - \max(0, H-s(\vec{d})) = f(\vec{d}) + s(\vec{d}) - 1$, which makes the same choices as Combine's $f(\vec{d}) + s(\vec{d})$ formula).  For intermediate values of the threshold ($0<H<1$), the average behavior of Penalize interpolates between these two extremes.

The Satisfice heuristic moves from being close to Maximize to achieving high social performance and low financial performance as the social threshold increases (i.e., as one moves from panel~(a) to panel~(c) in Figure~\ref{fig:performance-h}, Satisfice approaches the top-left corner).  This happens because the social threshold affects which performance dimension this heuristic will search.  When $H=0$, Satisfice will never switch to searching for social performance improvements, as the social threshold is obviously cleared.  Satisfice is then equivalent to Maximize.  Conversely, when $H=1$, the firm will only search on the social landscape, effectively turning Satisfice into a rule that could be called ``Maximize Social'' and which is the mirror image of Maximize but with respect to social performance (which is why Satisfice is close to the top-left corner in panel~(c)).  For intermediate values of the threshold ($0<H<1$), the average behavior of Satisfice interpolates between the two extremes.  Note that unless the social threshold is very high (as in panel~(c) in Figure~\ref{fig:performance-h}), Satisfice is usually dominated or closely matched by one of the other rules, as we see in all of Figure~\ref{fig:performance-k-rho} and in panels (a) and (b) in Figure~\ref{fig:performance-h}.

\subsection{Variability of Performance}

\begin{figure}[t]\centering
\includegraphics[width=\linewidth]{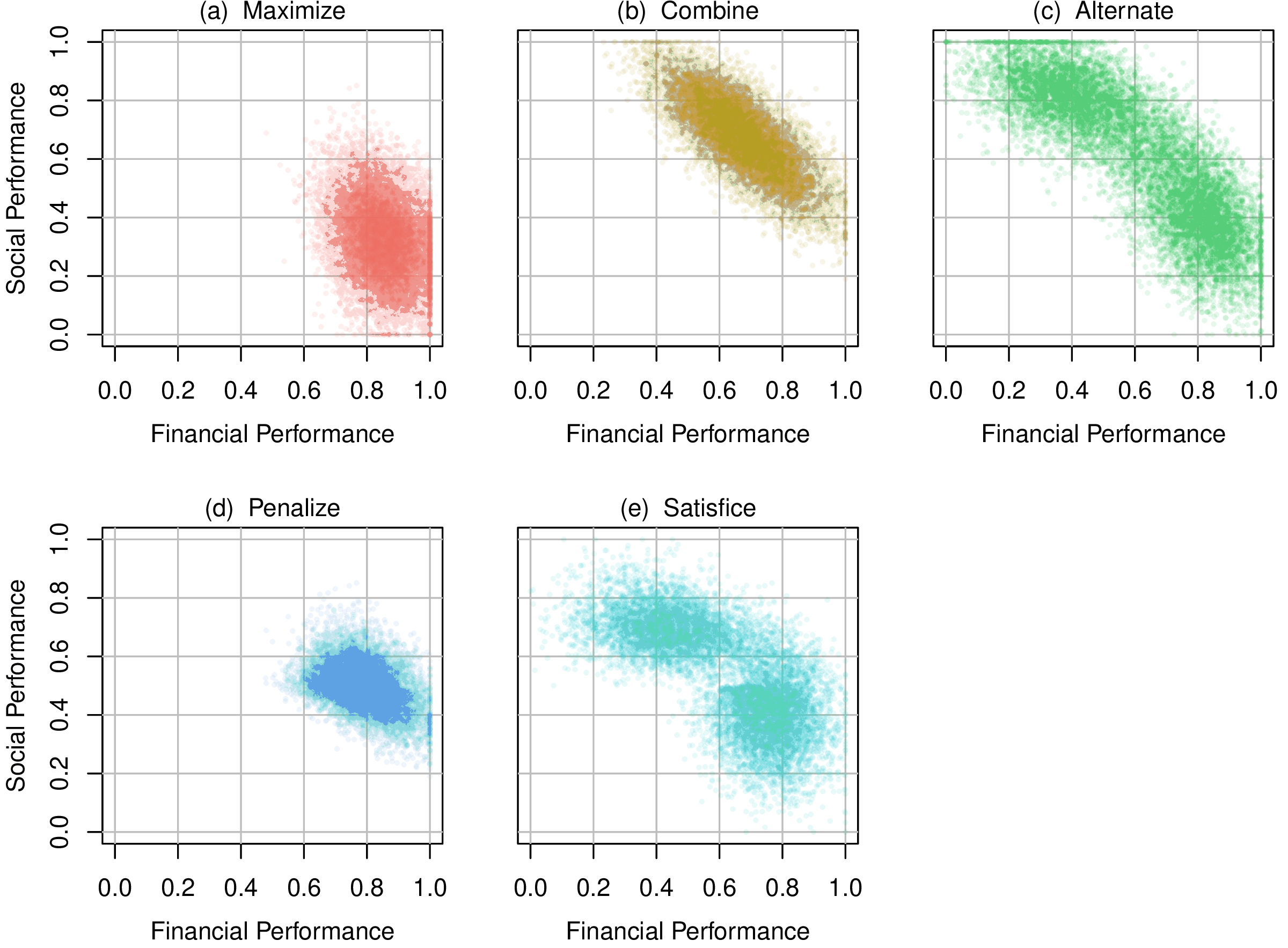}
\caption{Social and financial performance of 10,000 simulations of each heuristic.}\label{fig:performance-variability}
\end{figure}

So far, we have looked at the average performance of the five heuristics.  We now examine the variation around these averages; that is, we look at the whole range of performance for firms using a given heuristic.  This gives us an idea of the degree to which a heuristic determines firm performance.  Each scatterplot in Figure~\ref{fig:performance-variability} represents the social and financial performance at the end of the simulation ($T=200$) for 10,000 firms, each starting at a random position in a random landscape.  Each panel represents one heuristic.  The environmental parameters are kept fixed at $K=6$, $\rho=-0.5$, and $H=0.5$ (the same as in panel~(d) in Figure~\ref{fig:performance-k-rho}).

A general observation from Figure~\ref{fig:performance-variability} is that although within each panel there is much variability in the positions achieved (i.e., each cloud of points covers a large area), there are marked differences across panels (i.e., each cloud of points has a distinctive shape and location)---that is, heuristics have an important effect on which outcomes are more or less likely to occur.  Below, we examine what drives each panel's most salient characteristics.

The cloud of points representing Maximize is taller than it is wide (panel~(a)).  This happens because Maximize reliably gets firms to a local peak in the financial landscape while not paying any attention to social performance.  Hence, the width of the cloud of points in panel~(a) represents the range of financial local peaks, while the height of the cloud of points represents whatever social performance happens to be associated with those financial peaks.

The clouds of points for Combine and Alternate are diagonally oriented (panels (b) and (c)).  Even though the shape of the clouds is similar, the underlying reason is quite different.  Combine tries to find points that rank highly on the sum of both performance dimensions.  Because correlation is negative in Figure~\ref{fig:performance-variability}, high performance on one attribute will usually be associated with low performance on the other; hence the downward-sloping cloud.

The shape of the cloud for Alternate (panel~(c)) is due to the sequential nature of the search.  Recall that while this heuristic is searching on one dimension, the other dimension is ignored.  Hence, the cloud represents a snapshot of firms at different moments in their oscillation between goals: some are near the peak that Maximize would find and some are near the peak a ``Maximize Social'' heuristic would find.  Note that Alternate is more spread out than Combine.  This is because Alternate typically does not settle on a given point (unless that point happens to be a local peak in both landscapes).

Penalize (panel~(d)) produces the least variation in outcomes.  The reason is that Penalize ``sees'' a very rugged landscape, one that combines the ruggedness of the financial landscape and the ruggedness of the landscape representing the penalties (that is, Penalize sees the landscape generated by adding $f(\vec{d})$ and $-\max(0, H - s(\vec{d}))$).  Hence, firms using Penalize get stuck at a local peak more rapidly than with the other heuristics.

Satisfice (panel~(e)) leads to a bimodal distribution, as firms using this heuristic behave differently depending on whether or not they can clear the social threshold.  If they cannot clear it, they try to improve social performance until they do, at which point they try to improve their financial performance, and so on.  Thus, the bimodal distribution represents a search that oscillates between the points at which the two search modes are triggered.

One last observation from Figure~\ref{fig:performance-variability} is that the size of the clouds of points depends on the extent to which each heuristic explores the landscape.  That is, the heuristics that explore the least, like Maximize (which searches until getting stuck on a financial local peak) or like Penalize and Combine (which search on more rugged landscapes and, hence, get stuck earlier) exhibit less variability of outcomes than the heuristics that explore more, like Alternate and Satisfice, which switch between exploring the social and financial landscapes.  Appendix~\ref{app:random} shows that allowing occasional non-local (random) jumps does not alter these central qualitative conclusions, but as random jumps become more frequent, differences among heuristics compress, and outcomes drift toward chance-level performance.

\subsection{Outperforming Maximize}

Proponents of Maximize \citep[e.g.,][]{Frie70,Jens02} suggest that any multi-objective decision-making will fare worse financially than attempts to solely maximize financial performance.\footnote{A closely related critique in the CSR debate is that pursuing social objectives may distract from profitability and thereby reduce financial performance.  See, e.g., \citet{Karn11}.}  We investigate this statement by tracking the probability that our search heuristics perform at least as well as Maximize on \emph{both} performance dimensions.  When we say heuristic $A$ outperforms heuristic $B$, we mean that $A$ \emph{weakly dominates} $B$: $A$ achieves at least as high performance on both dimensions ($f_A \geq f_B$ and $s_A \geq s_B$) and may be strictly higher on one or both dimensions.  We use ``match'' to refer to the special case in which performance is equal on both dimensions.

\begin{figure}\centering
\includegraphics[width=\linewidth]{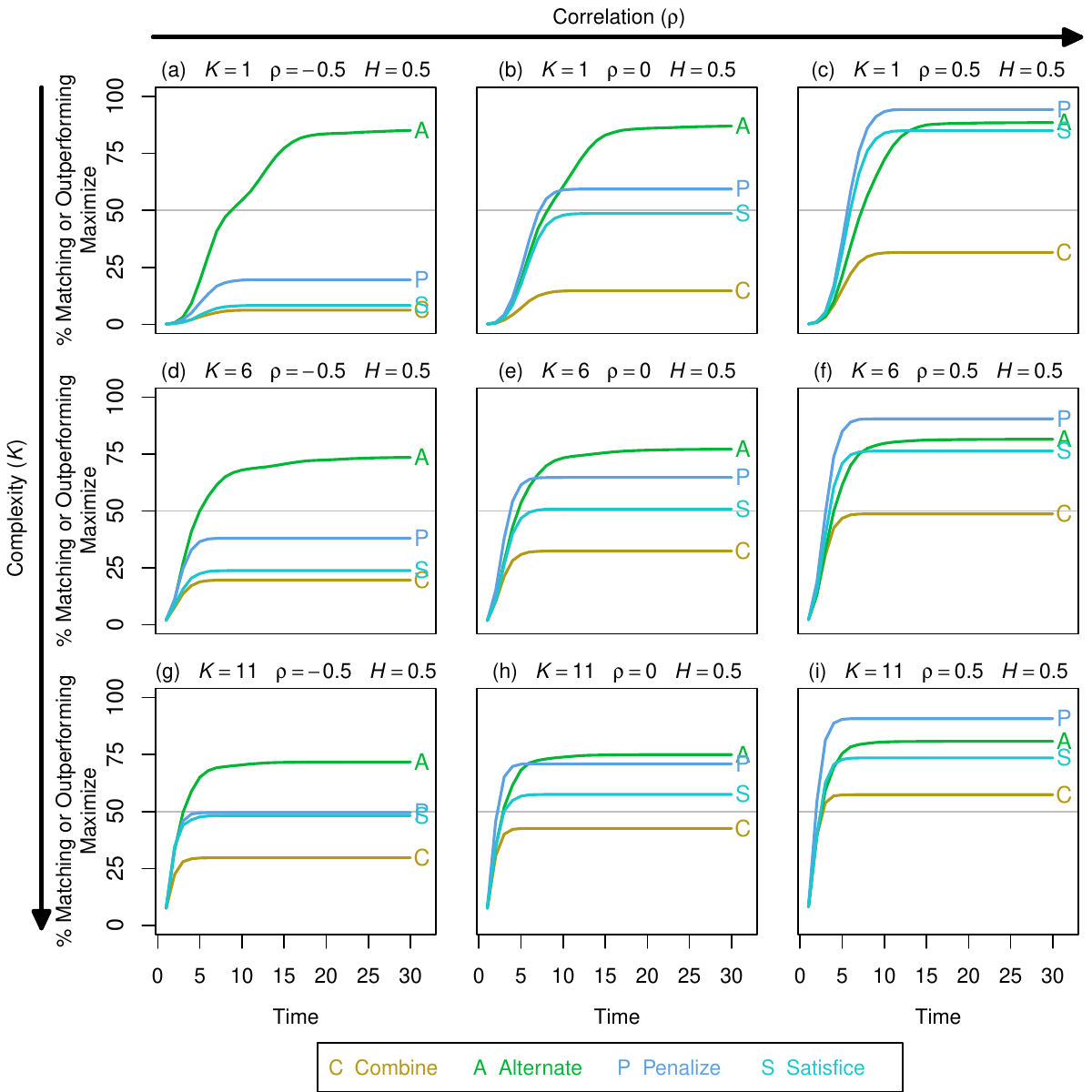}
\caption{Probability of matching or outperforming Maximize.}\label{fig:outp-maximize}
\end{figure}

Figure~\ref{fig:outp-maximize} plots the same nine scenarios as Figure~\ref{fig:performance-k-rho}; however, the $y$-axis now shows the cumulative probability of \emph{matching or outperforming} Maximize (that is, weakly dominating Maximize at some point during the search trajectory).  Specifically, at each point in time, we check whether each of the other four heuristics has, at any point up to that time, reached a position that weakly dominates Maximize's performance.\footnote{\raggedright More formally, the $y$-axes in Figure~\ref{fig:outp-maximize} measure performance for heuristic $H$ at time $T$ as $\frac{1}{\text{\#simulations}} \sum_\text{simulations} { \one \left[ \lor_{t=1}^T { \left( (f_t^H \geq f_T^\text{Maximize}) \land (s_t^H \geq s_T^\text{Maximize}) \right) } \right] }$.}  This measure counts the fraction of cases in which a heuristic that does not solely pursue financial gain can do at least as well as one that does.

Note that this comparison differs from Figure~\ref{fig:performance-k-rho} in an important way.  Figure~\ref{fig:performance-k-rho} reports where each heuristic is located at the end of the search process, on average.  Figure~\ref{fig:outp-maximize}, by contrast, asks whether a heuristic has discovered, at any point along its search trajectory, a strategy that weakly dominates the position reached by Maximize.  The figure therefore captures discovery rather than final location.  In practice, this assumes that firms can retain or revisit previously discovered alternatives, as when a firm explores multiple possibilities before settling on one encountered earlier.  For example, Microsoft's decision to rekindle its efforts to develop Windows after terminating its deep involvement with IBM's OS/2 \citep{Rich91} illustrates how firms may return to a previously explored trajectory after searching elsewhere.

This distinction explains why Combine can appear closer to the frontier in Figure~\ref{fig:performance-k-rho}, while Alternate stands out in Figure~\ref{fig:outp-maximize}.  Combine searches on the composite landscape $f(\vec d)+s(\vec d)$ and therefore tends to settle near balanced positions, but these need not match the financial local peak reached by Maximize.  Alternate, by contrast, may end farther from the frontier because it keeps shifting between objectives.  Yet that shifting also helps it discover new regions: after social search moves the firm away from Maximize's financial peak, later financial search can climb to a different financial peak.  Some of these peaks also have high social performance, allowing Alternate to weakly dominate Maximize in both dimensions.

The main result stemming from Figure~\ref{fig:outp-maximize} is a challenge to the proponents of Maximize: some heuristics match or outperform Maximize more than half of the time.  That is, some heuristics can often discover strategies that are at least as good as Maximize's current financial peak while also delivering at least as high social performance.  The explanation for this result is that bounded rationality dictates that firms cannot optimize, but only search.  This means that Maximize cannot really find the maximum financial performance but simply a local peak in the financial landscape.  In contrast, heuristics that explore more broadly can, given enough time, discover positions that dominate the particular local peak found by Maximize.  In other words, using heuristics that attend to multiple dimensions of performance can lead to \emph{both} better social performance and, in many cases, better financial performance than focusing narrowly on financial performance.

To get a better intuition for the underlying mechanisms driving this result, let us delve into some of the patterns observable in Figure~\ref{fig:outp-maximize}.  In all scenarios, Alternate is able to surpass the 50\% line (i.e., given enough time, Alternate matches or outperforms Maximize).  Reaching the 50\% line happens because half of the time Alternate starts its search on the financial landscape and therefore behaves exactly like Maximize until it reaches the financial local peak, after which it switches to searching on the social landscape.  Surpassing the 50\% line is due to Alternate's greater exploration: by alternating attention between the two landscapes, it can leave the financial local peak found by Maximize, move to a different region while searching socially, and then climb financially from that new starting point.  This process can reveal financial peaks that Maximize, because of its narrower path-dependent search, never visits.

Another observation is that when correlation is positive (i.e., panels (c), (f), and (i)), Penalize has the greatest probability of matching or outperforming Maximize.  This is because Penalize searches what can be termed a ``stretched financial landscape,'' where all points below $H$ have their financial performance reduced.  In these particular panels, financial outcomes exhibit positive correlations with social outcomes, meaning that this reduction of the landscape does not significantly alter the ranking of neighboring values.  Therefore, it is quite common for Penalize to locate the same position as that found by Maximize.

The finding that certain heuristics can identify strategies that match or surpass those of the Maximize approach suggests that these heuristics are effective at pursuing what \citet{Eno01} call an ``oblique strategy.''  By alternating attention between social and financial goals, heuristics like Alternate are able to zigzag across the performance landscape, exploring regions that Maximize, with its singular financial focus, typically overlooks.  This broader search allows Alternate to escape local financial optima and sometimes reach positions that deliver higher performance on both dimensions.  Thus, Alternate is not necessarily the best heuristic if performance is evaluated only by the average final position.  Its distinctive advantage is that it expands the opportunity set generated by search, making it more likely that the firm discovers a strategy that does as well as Maximize financially while also doing at least as well socially.

\section{Discussion}\label{sec:discussion}

A central challenge for boundedly rational organizations is how to search for strategies when performance is multidimensional and goals potentially conflict.  The behavioral theory of the firm has long recognized that organizations pursue multiple goals, yet the search literature has largely focused on single-objective landscapes.  Our study addresses this gap by examining how different search heuristics shape the discovery of alternative strategies, when firms must navigate dual performance dimensions.  Below, we elaborate on the broader contributions of our work.

\subsection{Contributions to Behavioral Strategy Research}

Our work contributes to understanding the relationship between multi-objective decision-making and organizational search.  The multi-objective decision-making literature has traditionally focused on how decision-makers evaluate and select among a given set of alternatives \citep{Keen76}.  This framing treats the set of alternatives as fixed and asks which should be chosen.  In contrast, the behavioral theory of the firm emphasizes that alternatives are not given but must be discovered through search \citep{Cyer63,Levi97}.  Our work bridges these perspectives by examining how evaluation heuristics shape the trajectory of organizational search---that is, how the criteria used to compare currently available alternatives influence which new alternatives the firm will subsequently discover.  In a multi-goal setting, this distinction is especially consequential because preferences are not only expressed in one-shot choice; they are implemented repeatedly, step by step, through a local evaluation rule that generates path dependence in what the organization ever learns is feasible.

This connection between evaluation and search helps explain why the Alternate heuristic can match or even surpass the performance of the single-goal focused Maximize approach.  Alternate does not outperform Maximize because it is superior at any single evaluation.  Rather, the sequential nature of its evaluation criteria produces a distinctive search trajectory: by oscillating between objectives, Alternate generates alternatives in regions of the landscape that Maximize, with its singular focus, is less likely to traverse.  The idea of oblique strategies captures this mechanism.  As Kay (\citeyear{Kay10}, Chapter 3) argues, firms may accomplish their objectives by not approaching them directly---that is, by avoiding narrowly focused and linear pursuit.  In our setting, the obliqueness stems from shifting the evaluative criterion over time: the wider pattern of alternative generation that emerges from Alternate's oscillating priorities enables it to uncover high-performing solutions that remain hidden to heuristics with more narrowly focused evaluation criteria.

Our results also add nuance to classic behavioral arguments about sequential attention to goals.  The behavioral theory of the firm has emphasized that firms may address conflicting goals by rotating attention among them (\citealt[\p 118]{Cyer63}, \citealt[\p 2553]{Fang18}, \citealt[\p 649]{Gaba19}).  In our simulations, however, simultaneous evaluation can, on average, outperform sequential evaluation (Combine outperforms Alternate throughout Figure~\ref{fig:performance-k-rho}).  Combine has in its favor the ability to continuously make trade-offs between dimensions, whereas Alternate's one-dimensional focus can drive it ``up'' one landscape at the expense of the other.  This is disadvantageous if the primary aim is to approach a balanced ideal point, but it can be advantageous when the aim is broader exploration---a distinction that becomes central when evaluating the probability of matching or outperforming Maximize.

Finally, we contribute by extending the NK landscape framework to study multiple performance dimensions.  While the NK framework is a staple in the search literature, it has customarily been used to examine search with respect to \emph{one} overall measure of performance, often called ``fitness.''\footnote{Although there is research in the NK literature that has modeled multiple sources of performance (e.g., \citealt{Csas16} assign performance to different attributes of a product and \citealt{Rivk03} and \citealt{Ethi09} assign performance to different divisions of a firm), such research has only studied how search heuristics affect just \emph{one} measure of fitness.} \citet{Adne14} developed a way to create landscapes with multiple performance dimensions, which they used to study spatial competition but not search.  Building on their work, we show how to model search on multi-attribute landscapes and how heuristic choice interacts with landscape structure (complexity $K$, correlation $\rho$, and threshold $H$) to shape \emph{joint} performance outcomes.  Our ``grammar'' for describing multi-objective heuristics (Figure~\ref{fig:heuristics}) provides a compact template that can be extended to additional rules in future work.

\subsection{Contribution to the CSR literature}

Jensen's \citeyearpar[\p 238]{Jens02} famous statement that ``[i]t is logically impossible to maximize in more than one dimension at the same time [\ldots]'' relies on a basic understanding of optimization.  We have suggested that the literature on multi-objective maximization provides a better theoretical foundation for reasoning about multiple performance dimensions.  This shift calls attention to the role of the search heuristics that take the place of the ``greater than'' operator used in unidimensional settings.  It also brings to the fore the distinction between preferences and heuristics and the fact that even heuristics that reflect the same preferences can generate different outcomes because they guide search differently.

Our work helps develop the role of decision processes in CSR research.  Prior research has often overlooked this dimension, in part because of the methodological challenges involved in studying how organizations actually make multi-objective decisions \citep[\p 953]{Agui12}.  The methodology we introduce allows us to open this black box by analyzing how alternative heuristic implementations translate into systematically different search trajectories and, consequently, different social and financial outcomes.

Because search heuristics are so consequential in the environments we model, choices should be made carefully.  In particular, some heuristics are rarely attractive: Satisfice is dominated (or closely matched) in most conditions, and Maximize systematically yields the lowest social performance, often leaving the firm outside the viable region implied by nontrivial social aspirations.  The practical implication is therefore not that one heuristic is universally best, but that heuristic choice can materially shape which compromises are reached and which opportunities are discovered under bounded search.

Choosing among heuristics is nuanced because heuristics can encode similar preferences while differing in implementation.  Penalize and Satisfice, for instance, both express contingent preferences---high financial performance subject to clearing a social threshold---yet Penalize is typically more effective at attaining that combination in our simulations.  Likewise, Combine and Alternate can both reflect joint concern for social and financial outcomes, but they operationalize that concern differently: Combine tends to move toward a balanced compromise and then settle, whereas Alternate tends to continue exploring by shifting attention across dimensions.  A key consequence of this exploration advantage is that Alternate can sometimes discover strategies that match or even outperform Maximize---including on financial performance---which is consistent with the view that pursuing social objectives can, under some conditions, open paths to superior financial outcomes \citep{Flam15,Doro17}.

Beyond explaining the workings of specific heuristics, developing the role of heuristics in CSR is valuable because it helps address the call by \citet[\p 31]{Marg09} to identify organizational practices that facilitate the joint achievement of social and financial objectives.  Our results also provide formal backing to research suggesting that valuable CSR initiatives can emerge from idiosyncratic search processes \citep{Haan13}, and they align with a central idea in stakeholder theory: that the role of management is to ``search for means to overcome trade-offs'' \citep[\p 331]{Hori14}.

We also clarify how environment structure affects the \emph{stakes} of heuristic choice.  Across the parameter ranges we examine, the relative ordering of heuristics is robust to changes in landscape complexity ($K$) and correlation between objectives ($\rho$), even though these contingencies shift absolute performance levels and can compress differences among heuristics (e.g., when $\rho$ is high, improvement on one dimension tends to carry over to the other).  Thus, while $K$ and $\rho$ shape how difficult joint improvement is, our results suggest that variation in how firms \emph{implement} preferences as search heuristics can be an important contributor to heterogeneity in social and financial outcomes---a possibility that complements the dispersion documented in the CSP--CFP literature \citep{Marg03,Orli03,Allo05,Wang16b,Gall19,Vish20}.

Our work also contributes to the literature on commensurability.  Prior research has emphasized that aggregating different performance dimensions---such as social and financial outcomes---often requires that these dimensions be made commensurable, or comparable on a single scale \citep[e.g.,][]{Carr99,Espe98}.  This requirement underpins the logic of heuristics like Combine and Penalize, which necessitate explicit weighting or translation of social and financial performance into a common metric.  However, achieving commensurability in practice is notoriously challenging, as the measurement and valuation of social outcomes are often contested and context-dependent \citep{Kroe14,Rawh17}.

Our findings suggest that strict cross-dimensional commensurability may be less essential for achieving balanced outcomes than previously assumed.  For instance, as shown in Figure~\ref{fig:performance-k-rho}, the expected performance of Alternate---a heuristic that does not require aggregating social and financial performance into a single metric---is remarkably similar to that of Combine, which does.  This does not mean that measurement of social performance is unnecessary.  Alternate still requires the organization to assess whether a move improves social performance when social performance is the focal dimension.  The point is narrower: Alternate does not require the organization to translate social performance into financial units, assign an exchange rate between the two dimensions, or evaluate marginal trade-offs between them on a common scale.

Put differently, organizations facing barriers to constructing a unified metric can sometimes approximate the outcomes of aggregation-based approaches by using heuristics grounded in sequential attention rather than simultaneous integration.  This resemblance is notable precisely because the heuristics impose different informational demands.  Combine and Penalize require cross-dimensional integration, such as weights or penalties that translate social performance into the same evaluative currency as financial performance.  Alternate requires within-dimension comparison: the organization must be able to assess whether a move improves social performance when social performance is the focal dimension, but it need not compare that improvement against a financial gain on a common scale.  A threshold-based heuristic like Satisfice can further reduce measurement demands by shifting the question from ``how much better?'' to ``have we cleared the threshold?''  This is particularly useful when it is difficult to assess marginal improvements along a social dimension \citep{Kroe14,Rawh17}.

More broadly, our approach shifts attention from whether performance dimensions can be perfectly commensurated to how organizations should select among heuristics with different informational and commensurability requirements.  Whereas prior work debates whether and how different types of value can be compared or combined, our results show that heuristics requiring cross-dimensional aggregation and heuristics relying on sequential attention or threshold checks can sometimes yield comparable outcomes.  This opens a wider practical menu for firms managing multiple objectives, especially when social performance can be assessed within its own domain but cannot be easily monetized or integrated into a unified metric.

\subsection{Contribution to Other Multi-Objective Contexts}

Our work provides an answer to Greve and Gaba's \citeyearpar[\p 323]{Grev20} call to develop a better understanding of search under multiple goals.  In particular, we analyze the trade-offs among a set of heuristics for pursuing multiple goals and show how heuristics that encode similar preferences can nonetheless generate systematically different search paths and outcomes.  Understanding such trade-offs may shed light on problems not only involving social and financial performance but also other (conflicting) goals such as willingness-to-pay and cost \citep{Port96}, short- and long-term performance \citep{Levi93}, and quality and quantity \citep{Shak82}.

Such multi-objective challenges are pervasive across individual and organizational contexts.  For instance, employee performance evaluations commonly require individuals to navigate incentive systems that combine multiple metrics---such as sales, customer satisfaction, and teamwork \citep{Kerr75,Ethi09}.  Firms also face multidimensional objectives when responding to industrial policy incentives that simultaneously target domestic investment, employment, and innovation outcomes \citep{Li19}.  In product development, design teams must iteratively search for configurations that balance competing criteria---performance, cost, time-to-market, and reliability---where each design decision (e.g., choice of materials, component architecture, or manufacturing process) jointly affects multiple outcomes \citep{Vino20}.  A team that evaluates prototypes solely on projected profit margin may converge on a local optimum that a team alternating attention between margin and user experience would have avoided, mirroring the dynamics our model captures.  In the healthcare sector, hospitals must balance service pricing decisions to ensure financial viability while simultaneously fulfilling population-health mission commitments \citep{Albe26}.  In all these settings, the problem of searching for satisfactory solutions under multiple goals is structurally similar to the one we have modeled, and our framework offers a generalizable approach for analyzing such decision processes.

Our framework also speaks to debates in the New Stakeholder Theory (NST), which examines how stakeholder engagement enables firms to create and appropriate value \citep{Barn18,McGa23,Kapl23}.  NST has been mostly silent, however, on how firms \emph{search} for the strategies that yield such value creation.  Its core normative question---how organizations should aggregate and prioritize heterogeneous stakeholder claims---can be recast in terms of the heuristics we formalize: some approaches rely on cross-dimensional aggregation and trade-offs, while others refuse such trade-offs through priority rules or aspiration-based thresholds, or shift attention sequentially across demands.  Yet as \citet{Lazz25} argues, NST has not sufficiently specified the normative assumptions underpinning its theoretical mechanisms, including how ``value'' is operationalized across stakeholder groups.  Our results add a behavioral dimension to this challenge: even when normative commitments are clear, the search heuristic used to implement them steers organizations toward different regions of the performance space.  Value creation is therefore not only about what firms \emph{aim} to create or how stakeholders define value, but also about which search procedures enable firms to \emph{discover} strategies that jointly improve multiple dimensions.

\subsection{Practical Implications}

Our research highlights two critical decisions firms must make.  First, they need to decide what position they want to occupy on the social--financial frontier.  The desired position will depend on managers' preferences as well as pressures stemming from competitors and stakeholders such as regulators, investors, and customers \citep{Batt22}.  Hence, different firms may aim for different targets.  For instance, a nonprofit may aim for point $Q$ in Figure~\ref{fig:sf-frontier} while a firm facing strong competitive and shareholder pressures may aim for point~$U$.

Second, once the desired location on the frontier has been determined, the firm needs to pick a heuristic likely to get it there.  Our research offers some initial guidance.  Maximize may be useful when financial performance is critical for survival and social performance requirements are undeveloped, as in nascent industries and markets that lack regulation and oversight.  Penalize is appropriate when financial performance is most important but clearing a self-imposed social threshold is desired.  Combine is most useful for firms that aim to jointly pursue social and financial goals and can sustain the commensurability infrastructure required to aggregate them.  Alternate is helpful when, apart from jointly pursuing both goals, the firm also wants to foster exploration---including exploring strategies that can ultimately exceed a purely financial searcher's local peak.  Satisfice could be chosen by firms aspiring to high social performance with only limited financial aims, as could be the case for nongovernmental organizations and certified benefit corporations \citep{Gehm19}, and it may be especially practical when fine-grained measurement of social performance is difficult.

We have also found that Figure~\ref{fig:sf-frontier}---the social--financial frontier---is a valuable framework in the MBA classroom, as it allows students to think more clearly about several issues related to CSR and multiple stakeholders: (i) that ``performance'' is not just financial performance, since all strategies also have a social performance; (ii) that trade-offs between social and financial goals are not always unavoidable---they are not when firms are inside the frontier or when new technologies push the frontier; (iii) that firms should aim to be on the frontier; (iv) that not all positions on the frontier are viable (only those that clear the social and financial thresholds); and (v) that picking a target position on the viable section of the frontier is an important choice that depends on the values of the firm's decision-makers.

\subsection{Future Work}

Our research can be extended in several ways.  In terms of empirical work, ethnographic or experimental research methods could be used to (i) surface additional heuristics that firms may use when dealing with multiple goals and (ii) identify the situations in which the different heuristics are more likely to be used.  Future empirical work could also test the predictions of our model using a large sample.  One way would be to use the composition of the board of directors or the top management team (TMT) as a proxy for the type of heuristics used by the firm.  For instance, one could test whether firms with finance-heavy representation in the TMT tend to perform like the Maximize heuristic or whether firms with high-turnover TMTs tend to perform like the Alternate heuristic.  The social threshold could be evaluated by measuring the extent to which a company's ESG goals exceed regulatory standards.  Future work could examine the implementation costs of multi-objective heuristics, including their measurement demands.

In terms of theoretical work, future research could (i) study heuristics that deal with more than two performance dimensions, as in designing a product to satisfy many different stakeholders, (ii) study heuristics that include multiple actors, some of whom may have opposing ideas regarding the firm's proper goals, and (iii) incorporate processes by which firms may actively shape how performance is incentivized and measured.  These extensions would bring the model closer to multi-stakeholder settings by decomposing ``social performance'' into multiple stakeholder-specific dimensions and allowing goal conflict to shape search.

Related to this last point, a promising direction for future research is to connect our findings to the emerging literature on market shaping \citep{Helf21,Vino19} by considering how the adoption of particular search heuristics can itself serve as a lever for shaping markets.  In multidimensional contexts, the way firms ``see'' the landscape is mediated by their chosen heuristic---for example, the Combine heuristic aggregates social and financial dimensions, while the Alternate heuristic focuses on one dimension at a time.  This implies that shaping is not limited to directly changing the social and financial payoff structures, but can also involve influencing which heuristics firms adopt, whether through regulation, industry standards, or stakeholder pressure.  For instance, policy interventions might mandate that firms apply specific weights to different objectives, effectively institutionalizing a preferred heuristic and thereby guiding organizational search and aggregate market outcomes.  Exploring how such mechanisms affect the distribution of firms along the social--financial frontier, and the collective achievement of goals like sustainability, offers a fruitful avenue for extending both shaping theory and the study of multi-objective decision processes.

\subsection{Conclusion}

This paper sheds light on the fundamental behavioral challenge of how organizations search for satisficing strategies when multiple objectives matter.  We show that some multi-objective search heuristics are much more likely than others to achieve particular stakeholder goals.  Choosing a search heuristic is therefore a critical matter.  Interestingly, this choice depends mostly on what the firm wants to achieve rather than on characteristics of its environment.  To develop these insights, we have extended the search literature to formally study search under multiple goals.  We have thus contributed to opening the black box of search strategies for stakeholder value while---in the manner of some of the heuristics we model---pursuing the dual goals of rigor and relevance.

\clearpage
\begin{singlespacing}
\renewcommand{\refname}{References{\vskip -5pt}}
\bibliography{bibliography}
\end{singlespacing}

\clearpage
\appendix
\FloatBarrier
\section{The Weighted Combine Heuristic}\label{app:weighted-combine}

The Weighted Combine heuristic, denoted $\mathrm{Combine}(w)$, searches on the combined landscape $w f(\vec{d}) + (1-w) s(\vec{d})$.  For example, Combine(0.8) assigns 80\% weight to financial performance and 20\% to social performance.  Combine(0.5) is equivalent to the Combine heuristic analyzed in the body of the paper.

\begin{figure}[h]\centering
\includegraphics[width=\linewidth]{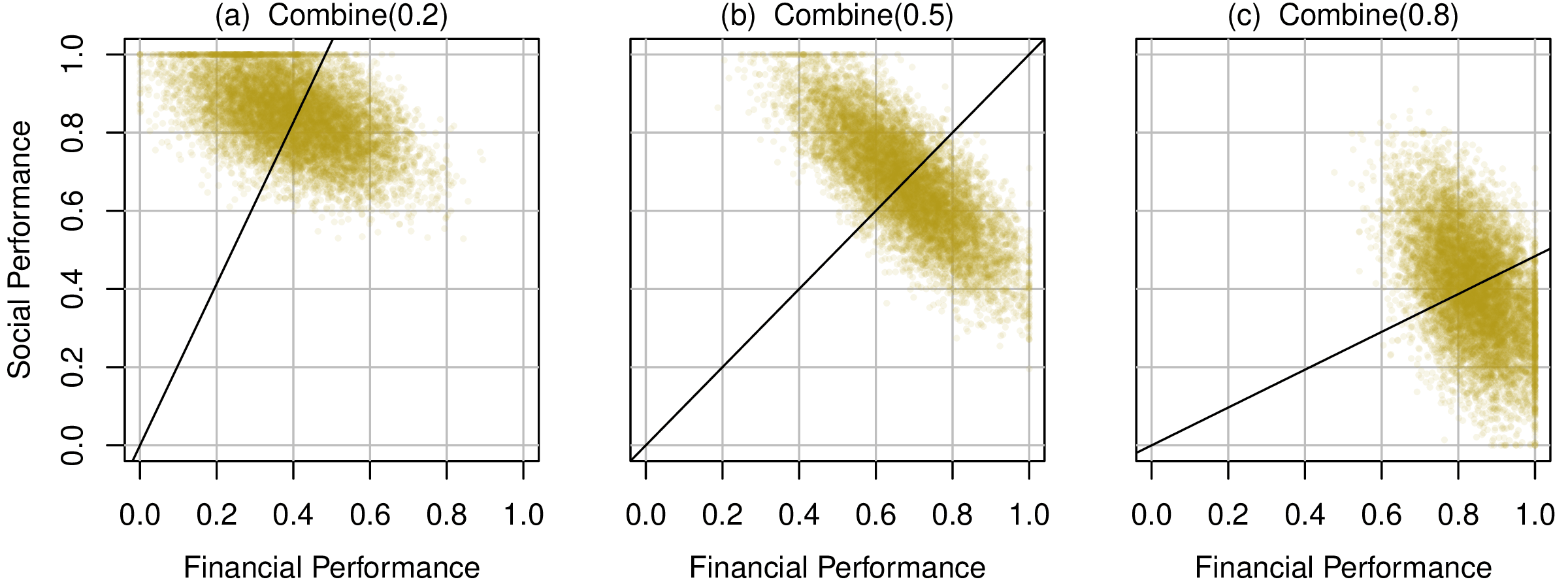}
\caption{Illustration of the Weighted Combine heuristic under the same conditions as Figure~\ref{fig:performance-variability}.}\label{fig:weighted-combine}
\end{figure}

Figure~\ref{fig:weighted-combine} simulates 10,000 runs of $\mathrm{Combine}(w)$ for $w=0.2$, 0.5, and 0.8 under the same assumptions used to generate Figure~\ref{fig:performance-variability}.  The main effect of $w$ is to tilt outcomes toward the weighted dimension: as $w$ increases from 0 to 1, search shifts from high-social/low-financial performance toward low-social/high-financial performance.

\section{Relaxing the Local Search Assumption}\label{app:random}

A key assumption in our baseline model is strict local search: firms only evaluate strategies that differ by a one-bit change.  To examine robustness to this assumption, we allow firms to follow their focal heuristic most of the time but, with a small probability, make a non-local (random) move.\footnote{Conceptually, such non-local moves capture episodes of more exploratory variation (e.g., experimentation with distant projects, acquisitions, or major strategic reorientations) that can arise under conditions such as crisis, ownership change, or active investor intervention.  This is consistent with classic arguments that organizational adaptation reflects a shifting balance between exploration and exploitation \citep[see, e.g.,][]{Marc91}.}  Occasional non-local moves reduce the extent to which any one heuristic becomes ``trapped'' at a particular local peak, thereby dampening (though not eliminating) differences that arise purely from locality constraints.

\begin{figure}\centering
\includegraphics[width=\linewidth]{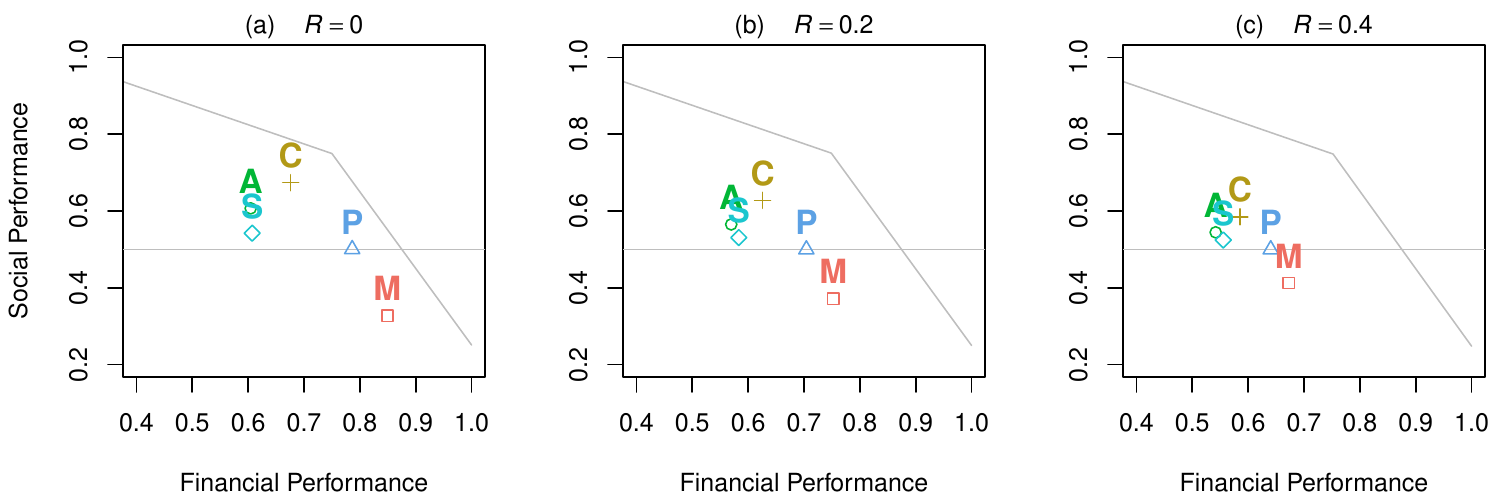}
\caption{Performance plots for different exploration probabilities ($R$).}\label{fig:random}
\end{figure}

We model exploration via a parameter $R$, which denotes the probability that the firm performs a ``random jump'' rather than a step of local search in a given time period.  Thus, $R=0$ corresponds to the analyses reported in the body of the paper (i.e., full local search), while, for example, $R=0.2$ corresponds to a 20\% probability of performing a random jump per period.  A random jump is modeled as drawing a new decision vector by assigning each of the $N$ binary decisions in $\vec{d}$ at random (consistent with \citealt[\p 938]{Levi97}).  Figure~\ref{fig:random} uses the same $K$, $\rho$, and $H$ values as panel~(d) in Figure~\ref{fig:performance-k-rho} while varying the probability of performing a random jump ($R=0$, 0.2, and~0.4).

Several patterns emerge from Figure~\ref{fig:random}.  First, the qualitative ranking of heuristics is preserved across all three panels: Maximize attains the highest financial performance but the lowest social performance, while Combine and Alternate remain the strongest on the social dimension.  Second, Maximize falls below the social threshold $H=0.5$ in every panel, indicating that random jumps alone do not remedy the social-performance deficit of a purely financial focus.  Third, the spread among heuristics shrinks as $R$ grows: random jumps dilute each heuristic's local search logic and compress outcomes toward one another (i.e., closer to random performance $(0.5,0.5)$).  At $R=0.4$, Combine, Alternate, and Satisfice are nearly indistinguishable, since their distinct local search patterns are increasingly overridden by random exploration.  Overall, relaxing strict local search dampens but does not eliminate the differences among heuristics.

\end{document}